\documentclass[preprint,5p,times,twocolumn]{elsarticle}

\usepackage{amssymb}
\usepackage{amsmath}
\usepackage{graphicx}
\usepackage{hyperref}
\usepackage{cleveref}
\usepackage{tabu} 
\usepackage{tabularx}
\usepackage{booktabs}                  
\usepackage{mathptmx}                  
\usepackage{multirow}
\usepackage{color}
\usepackage{enumitem} 
\usepackage{color}
\usepackage{subfig}
\usepackage{subcaption}  
\usepackage{float}       
\usepackage{placeins}    

\newcommand{\rev}[1]{{#1}}

\journal{Computers \& Graphics}

\begin{document}

\begin{frontmatter}



\title{Enhance Comprehension of Over-the-Counter Drug Instructions for the General Public and Medical Professionals through Visualization Design}

\author[label1,label2]{Mengjie Fan}
\author[label3]{Katrin Angerbauer}
\author[label4]{Yinchu Cheng}
\author[label4]{Yingying Yan}
\author[label4]{Xiaohan Xu}
\author[label5]{Tianfu Wang}
\author[label3]{Michael Sedlmair}
\author[label2]{Yu Yang*}
\author[label1,label2]{Liang Zhou*}

\affiliation[label1]{organization={Institute of Medical Technology, Peking University Health Science Center},
            city={Beijing},
            postcode={100191},
            country={China}}

\affiliation[label2]{organization={National Institute of Health Data Science, Peking University},
            city={Beijing},
            postcode={100191},
            country={China}}
\affiliation[label3]{organization={Visualization Research Center (VISUS), University of Stuttgart},
            city={Stuttgart},
            postcode={70569},
            country={Germany}}
\affiliation[label4]{organization={Peking University Third Hospital},
            city={Beijing},
            postcode={100191},
            country={China}}
\affiliation[label5]{organization={School of Journalism and Communication, Peking University},
            city={Beijing},
            postcode={100871},
            country={China}}
\affiliation{organization={*: equal contribution}, country={corresponding authors}}

\begin{abstract}
  Drug instructions are crucial for guiding the rational use of medication. 
  We conduct a visualization design study to enhance the comprehension of over-the-counter (OTC) drug instructions, targeting both the general public and medical professionals. 
  We devise two tailored drug instruction designs for different audience groups through an iterative design process. 
  A controlled user study reveals that our design outperforms traditional text-based instructions in terms of response time and usability, and the availability of two versions is also found to be beneficial. 
  This study also motivates a taxonomy based on a systematic classification of OTC drug instructions sampled from an official drug database, which received positive expert feedback. 
  Finally, this study summarizes a workflow for a visualization design strategy based on our design exploration and user study feedback, which can be generalized to other OTC drug instructions.
\end{abstract}


\end{frontmatter}



\section{Introduction}
\label{sec1}
Drug instructions, approved by regulatory authorities such as the U.S. FDA, European Medicines Agency (EMA), and China National Medical Products Administration (NMPA), serve as a critical foundation for ensuring patient safety and guiding appropriate medication use~\cite{who2024}.
By helping people quickly and accurately understand drug instructions, we can reduce irrational medication use and enhance patient safety. 

Drug instructions consist of various parts containing different types of data and typically involve the general public (including patients and those who lend help with their medicine) and medical professionals, who have varying information needs and medical knowledge regarding a specific drug. 
These characteristics indicate that conveying drug instructions is a ``wicked problem'' similar to patient data communication~\cite{Rajabiyazdi2021} that requires understanding the different needs of different data for different stakeholders (e.g., patients, doctors) and, therefore, is challenging. 

Current drug instructions often fail to meet patients' needs for practical and comprehensible information ~\cite{Raynor2007a} due to several challenges.
First, small and dense text layouts often make reading and understanding difficult, particularly for older adults.
Second, the excessive use of medical terminologies poses significant barriers for nonprofessionals, reducing their ability to comprehend essential information~\cite{Martin2005}. 
Lastly, variations in drug usage and dosage across patient groups and indications add complexity, increasing the risk of misinterpretation and potential misuse.

Visualization has the potential to enhance the design of drug instructions by simplifying complex data~\cite{Abudiyab2022, Gotz2016}, reducing information overload~\cite{Caban2015}, and supporting intuitive decision-making and analysis~\cite{Gotz2016, Caban2015, Chishtie2022}.
As a crucial form of medical communication~\cite{Manchanayake2018}, drug instructions can benefit from visual elements. 
Visualizations have already proven effective in patient education~\cite{Meuschke2022} and in facilitating doctor-patient communication on various diseases and medical procedures~\cite{Hakone2017}. 
By integrating visualization into drug instructions, we may improve the clarity~\cite{Abudiyab2022}, understanding~\cite{Caban2015}, and overall effectiveness~\cite{Gotz2016} in conveying essential information to related stakeholders.
However, interactive visualizations tailored to the specific content of drug instructions, which aim at improving understanding and communication across different populations, are lacking.

\rev{
Therefore, the primary goal of this paper is to develop and evaluate an interactive visualization framework for over‑the‑counter (OTC) drug instructions to improve comprehension for diverse stakeholders. This is crucial as it directly addresses the need for customized and interactive visual designs for drug instructions. Successfully realizing this goal can mitigate the known issues of textual instructions, thereby enhancing medication safety and communication efficacy.}


To achieve this goal, we conduct a design study with pharmacists ($N$=3) and iteratively develop a prototype of OTC drug instructions for different stakeholders over a 12‑month period. 
This process includes designing and implementing interactive visualization frameworks for several commonly used OTC drugs with different purposes. 
Based on lessons from a failed prototype, we identify key design factors for visualizing OTC drug instructions. 
We systematically review the NMPA database of OTC drugs, generalize design patterns, and propose a taxonomy as shown in~\Cref{fig:design_otc}(1).
Our prototype is finalized after two all-hands meetings with the pharmacists and numerous minor feedback in between.

The effectiveness and acceptance of the proposed visualization drug instructions are evaluated.
The final prototype of a specific OTC drug instruction is evaluated with potential target users from the general public in a controlled user study ($N$=60). 
The results demonstrate that our prototype has a shorter response time and higher usability rating compared to traditional text-based instructions.
A design workflow (\Cref{fig:design_otc}) is then proposed based on the design study and user study.
The workflow outlines how to develop a visualization framework and design each part of OTC drug instructions.

In summary, the main contributions of our work are as follows.
\begin{itemize}  
    \item A design study of a visualization framework for OTC drugs, resulting in a prototype for different target users, as well as a taxonomy and workflow for visualizing OTC drug instructions.
    \item A user study evaluating the effectiveness of a visualization version of a frequently-used OTC drug instruction against the traditional text version. 
\end{itemize}

To the best of our knowledge, our work is the first attempt at systematic visualization design for drug instructions as a whole, considering different populations. 
Furthermore, our visualization design, taxonomy, and workflow are generic and can be potentially adapted to drug instructions worldwide.

\section{Pharmaceutical Background}
\label{sec: background}

Irrational use of medication is a significant global issue~\cite{Ofori-Asenso2016} that may lead to consequences including misuse, inappropriate dosing (overdose or underdose), drug addiction, and adverse drug events~\cite{Melku2021}. 
According to the World Health Organization (WHO), 
ensuring the rational use of medicines requires that ``patients receive appropriate medications that meet their clinical needs and are prescribed at the correct dosage''~\cite{who2024medicine}.

To ensure patient safety, drugs are classified as prescription or over-the-counter (OTC) based on usage, safety, dosage, specifications, and routes of administration~\cite{Leelavanich2020}. 
Prescription drugs require a licensed physician or authorized medical professional to prescribe them and must be used under their supervision. In contrast, OTC drugs can be purchased and used by consumers without a prescription, as they are intended for conditions that patients can self-identify and manage~\cite{Chang2016, Ray2022}.
Some medications, known as ``dual-status drugs'', can be used as either prescription or OTC depending on their usage and dosage.
In this work, we focus solely on visualizing OTC drug instructions.

OTC drugs have a long history of clinical use~\cite{Aronson2004}, well-established efficacy, and are generally considered safe~\cite{Bukic2019}. However, prolonged or excessive use can still lead to adverse reactions~\cite{Ray2022, Tanibuchi2024}. It is important to note that such reactions can vary significantly from person to person~\cite{Alomar2014}. Therefore, it is crucial to follow the instructions carefully when using OTC drugs and to monitor for any adverse effects after taking them.

Drug instructions are generic in many countries, see examples of instructions for paracetamol/acetaminophen in different countries in~\Cref{fig:similar_structure}.
We also provide the link or PDF of these three drug instructions in the supplemental material (1.supplemental\_material\_main.pdf, 3. labeled\_instruction\_paracetamol\_chinese.pdf, 4. Paracetamol665mgTab\_english.pdf).
Drug instructions mainly include drug names, indications, usage and dosage, contraindications, precautions, adverse reactions, drug interactions, and preservation conditions, among other things. 
The contents related to drug use should be carefully read before taking the medication to avoid hidden risks.

\begin{figure}[!htb]
  \centering  
  \subfloat[China]{\includegraphics[width=0.8\linewidth]{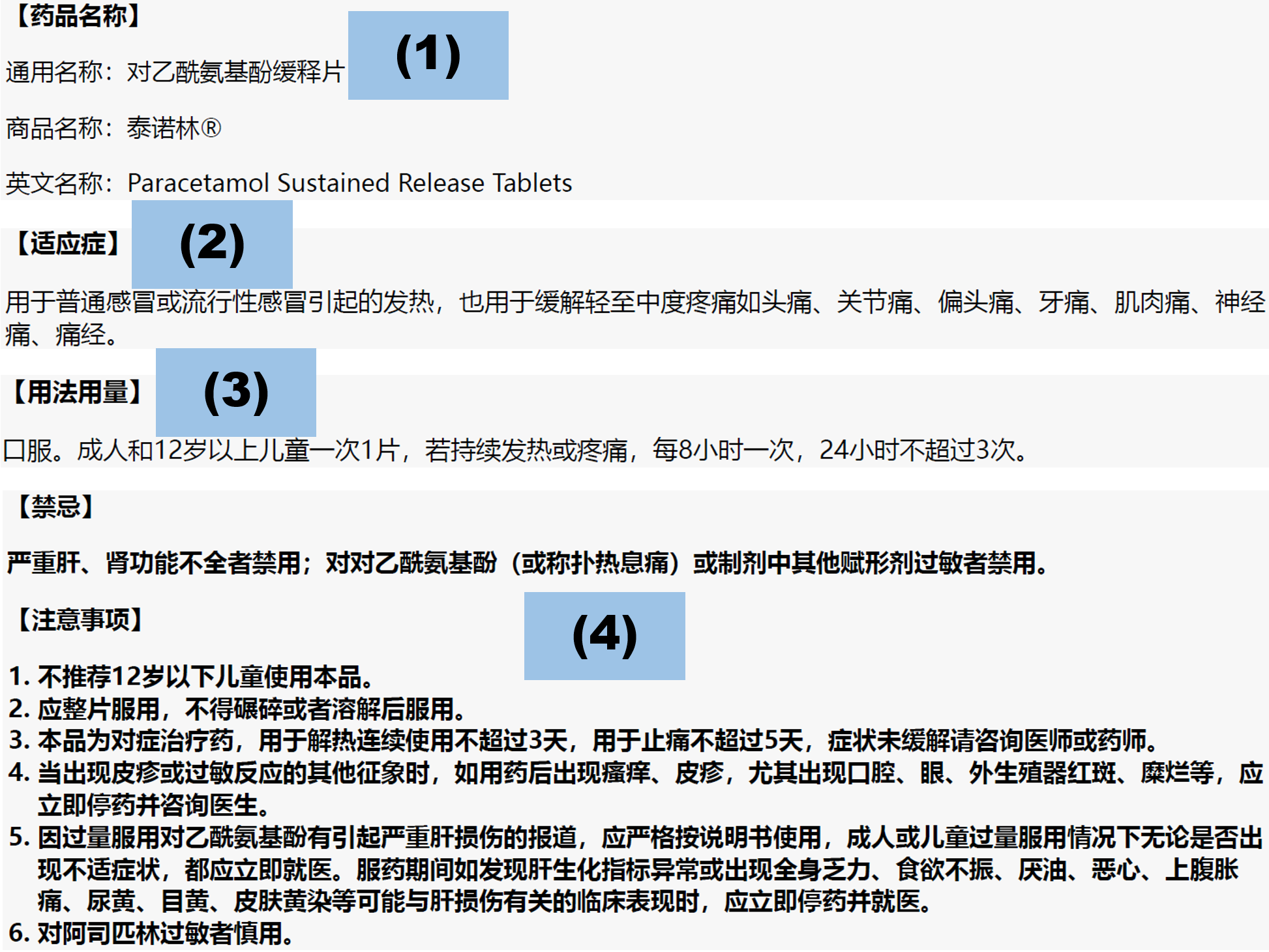}\label{fig:china_version}}\\
  \subfloat[US]{\includegraphics[width=0.8\linewidth]{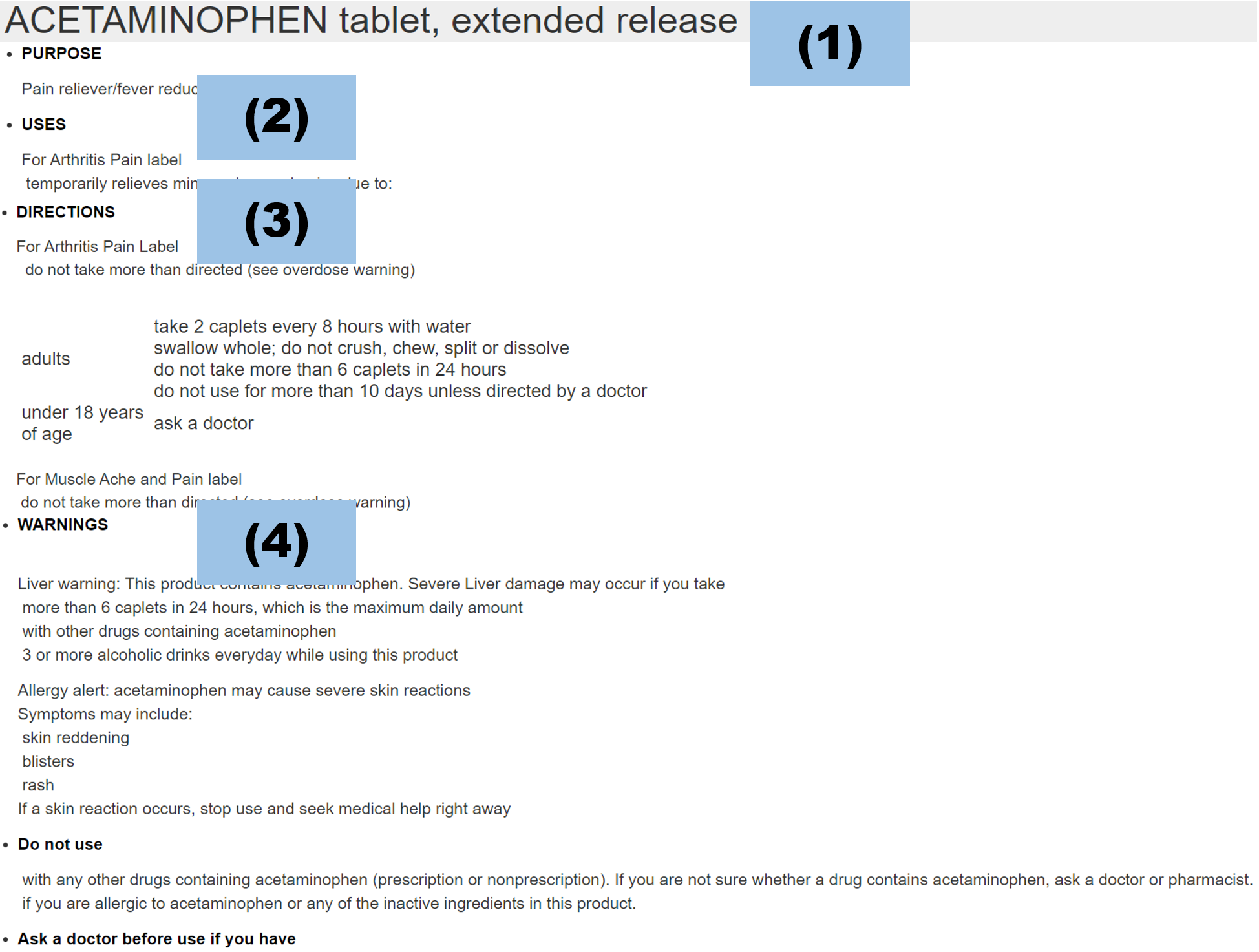}\label{fig:US_version}}\\
  \subfloat[UK]{\includegraphics[width=0.8\linewidth]{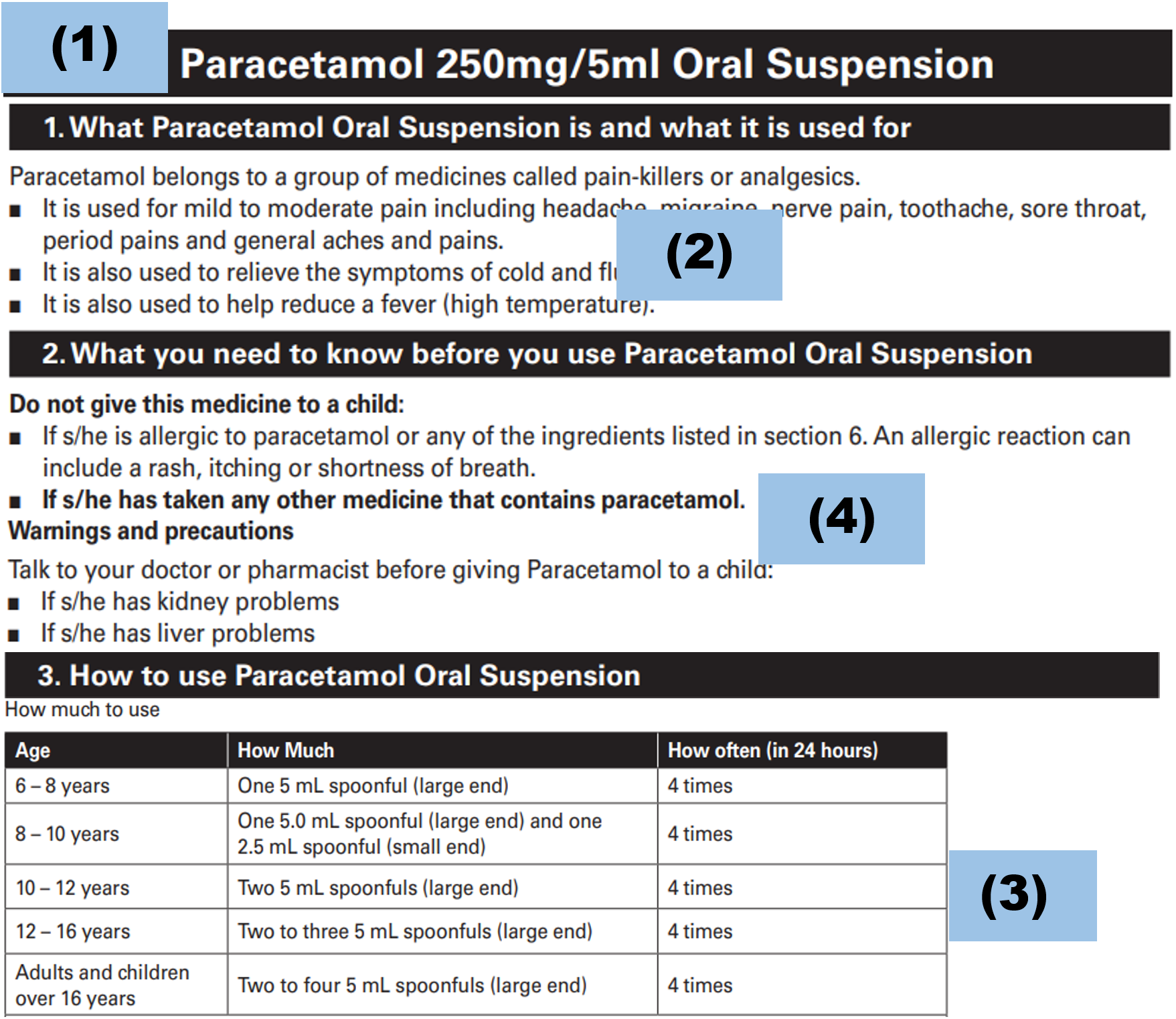}\label{fig:UK_version}}  
  \caption{Drug instructions for paracetamol/acetaminophen in China (a), US (b), and UK (c) are shown here. Drug instructions typically consist of (1) drug name, (2) indications, (3) usage and dosage, and (4) warnings and precautions, etc.}
  \label{fig:similar_structure}
\end{figure}

However, existing drug instructions are often too complex for users with limited medical knowledge to understand thoroughly and efficiently~\cite{Martin2005}. 
Understanding drug instructions can be especially challenging for the elderly~\cite{Williams1995}.
To address this, a pilot project plan~\cite{cde2024drug}
that includes guidelines for creating \emph{the simplified version} and the \emph{large print version} of drug instructions, as well as the electronic version (\emph{complete version}) was issued. 
Similarly, the Patient Medication Information (PMI) Proposed Rule~\cite{FDA2024PMI}
was proposed to ensure that patients receive crucial information about their medications through simplified, standardized, and accessible formats, improving patient understanding and adherence.
These initiatives highlight the urgent need for designing more effective and patient-centered drug instruction formats.  

\section{Related Work}
Our study combines three research areas: pharmaceutical information comprehension, visualization for medical communication, and visualization design studies.
The following gives an overview of recent works in these areas.

\subsection{Pharmaceutical Information Comprehension}
\label{pharma_comprehension}
Researchers explore the specific support people need to make sense of visualizations, aiming to improve visualization design~\cite{Rezaie2024}, which may enhance comprehension.
Pictographic visualizations could improve user engagement and comprehension~\cite{Haroz2015, Burns2022}.
A systematic review~\cite{Schubbe2020} concludes that it is beneficial to include pictures in medical communication, and icons/pictograms with less text may be most helpful. 

Many early studies regarding pharmaceutical information comprehension focus on pictograms/icons like the United States Pharmacopeial Convention (USP)~\cite{USP2024pictograms},
the International Pharmaceutical Federation (FIP)~\cite{fip2024pictograms},
and the Risk/Benefit Assessment of Drug Analysis and Response (RAD-AR)~\cite{RAD-AR2024pictograms}
Council of Japan~\cite{Kanji2018, Abdu-Aguye2023}, which are corroborated to help reduce the reading burden of patients, facilitate comprehension, and improve medicine adherence~\cite{Houts2006, Sharif2014, Ng2017, Merks2021, Gutierrez2022}.
Adding pharmaceutical pictograms to written text significantly improves medication understanding among older adults~\cite{Ng2017}.
Therefore, including pictograms on medication labels is recommended to enhance communication. 
While symbols boost comprehension, studies suggest that symbol-plus-text and text-only formats outperform symbol-only formats~\cite{Mayhorn2009}, indicating that using only pictograms may not be ideal.
Among older Singaporeans, pictograms for dosage and precautions are more effective than those for indications~\cite{Malhotra2023}.
A pictogram storyboard~\cite{who_fip2024} was designed to convey medication information but offers limited details, requiring users to click to access specific pictograms or manually indicate the medication's site of action on a body outline.

Several studies have explored how to enhance comprehension of numerical information of drug instructions~\cite{Sharko2022} and designed visualizations like ``liquid dosing''~\cite{Yin2017}, ``maximum dosing''~\cite{King2011}, and ``time interval''~\cite{Morrow2004}.

\rev{Collectively, these studies provide strong evidence for the value of visual elements in improving specific aspects of medication communication. However, the focus has largely been on enhancing comprehension of isolated content types (e.g., dosage, timing, or specific risks) or on evaluating individual pictograms. A significant gap remains in the systematic, holistic visualization design for the entire drug instruction document, which requires to integrate diverse data types (textual, numerical, categorical) and prioritize information for different stakeholders.}

\rev{To fill this gap, our work aims to move beyond isolated pictograms or charts. We propose a comprehensive visualization framework that designs the overall structure of drug instructions, highlighting critical information holistically and selecting appropriate visual forms (pictograms, charts, interactive elements) based on the specific content and context.}

\subsection{Visualization for Medical Communication}
In current data-driven medical practice, medical visualization is a key research area, focusing on access to medical evidence and helping stakeholders understand health data~\cite{Wang2019}. 
A recent review~\cite{Ma2024} summarizes four applied domains of visualization for narrative medicine, one of which is medical communication. 

Researchers have explored various methods to enhance communication between medical professionals, patients, and other stakeholders through visual tools~\cite{Abudiyab2022}.
Lifelines~\cite{Plaisant2003}, a classical medical information visualization method, supports the representation of diagnoses, test results, or medications, giving intuitive access to data comprehension.
PROACT, a patient-centered visualization tool, effectively communicates prostate cancer risk using simple chart visualizations and an AI-assisted predictive model~\cite{Hakone2017}.
Interactive dashboards, using daily living and patient-reported outcomes data, are designed to improve pediatric care~\cite{Lakshmi2018}.
A narrative visualization that integrates medical conditions can effectively enhance communication through analysis of data, including prescription data~\cite{So2021}.
By breaking down disease data and using medical visualization with narratives, data-driven disease stories become accessible to the public~\cite{Meuschke2022}.

\rev{The aforementioned studies demonstrate the successful application of visualization for communicating complex medical data, primarily focusing on disease states, health trends, or treatment outcomes for specific populations. While these approaches share the goal of improving understanding, the domain of drug instructions presents a distinct and highly structured communication challenge. It requires conveying standardized, regulatory-mandated information to a spectrum of users with varying literacy and expertise, where misunderstanding carries immediate safety risks.}

\rev{Consequently, visualization techniques designed for narrative health data or specific diseases are not directly transferable. There is a lack of systematic visualization research targeting the drug instruction document as a primary object of study. Our work directly addresses this gap by investigating how to visualize the integral components of drug instructions, with the specific goal of enhancing safe and effective use for the general public and medical professionals.}

\subsection{Visualization Design Studies}
A design study is a project where visualization researchers address the specific real-world problem encountered by domain experts~\cite{Sedlmair2012}. 
Practical guidance for design studies includes categorizing visualization methods by task and data type~\cite{Shneiderman1996}, using problem-driven and technique-driven approaches~\cite{McKenna2014}, applying a nested model for design and validation~\cite{Munzner2009}, following a nine-stage framework~\cite{Sedlmair2012}, hosting creative visualization-opportunities workshops~\cite{Kerzner2019}, using VizItCards for high-quality designs~\cite{He2017}, and utilizing the Five-Design Sheets (FDS) method for iterative design~\cite{Roberts2016}. Additionally, six criteria have been established for rigorous visualization design studies~\cite{Meyer2020}.
Considering the different types of data, we adopt a problem-driven iterative design process in this work.

Design studies are mainly user-centered designs tailored to specific domain problems, such as identifying unknown errors in engine manufacturing~\cite{Eirich2022}, managing emergencies~\cite{Liu2008}, detecting network threats~\cite{Chen2014}, analyzing genomic epidemiology~\cite{Crisan2022}, designing customizable food trackers~\cite{Luo2019}, and addressing health data discrepancies~\cite{Mccurdy2019}. Our work also contributes to the health domain, but with a focus on enhancing health data communication.

Researchers have recognized the importance of considering individual and group differences in data visualization systems rather than relying on ``one-size-fits-all" interfaces~\cite{Liu2020}. 
Wicked problems, where stakeholders disagree on problem definitions and solutions~\cite{Rajabiyazdi2021}, pose a challenge for designing visualizations, especially when addressing complex issues involving diverse groups of people with varying expertise, background, needs, and perspectives~\cite{Mahyar2020, Burns2023}.
In medical contexts, this complexity is heightened by the different perceptions and needs of patients, doctors, nurses, pharmacists, etc.
To create effective visual drug instructions, it is essential to understand and address the specific needs of both medical professionals and nonprofessionals. 

\rev{This body of work establishes that effective visualization design for complex domains should be problem-driven and account for diverse stakeholders---a paradigm well-suited to the challenge of drug instructions. As noted previously, communicating drug instructions is a ``wicked problem'' with conflicting stakeholder needs. Therefore, a visualization design study that actively engages these different stakeholders (e.g., pharmacists and the general public) is not merely appropriate but necessary to navigate this complexity.}

\rev{Our study is positioned within this methodological framework. We conduct a design study focused specifically on designing a visualization system for drug instructions, directly engaging representative stakeholder groups (pharmacists and the public) to ensure the design addresses their distinct and often competing needs, thereby providing a structured response to this wicked problem.}

\section{Visualization Design Study for Drug Instructions}
\label{sec: visualization_design_sec}
\begin{figure}[ht]
  \centering
  \includegraphics[width=\linewidth]{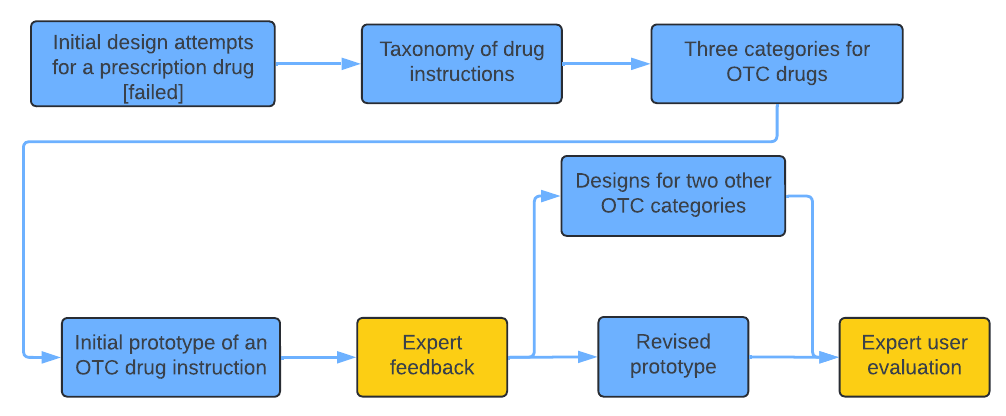}
  \caption{The process of our visualization design study for drug instructions.}
  \label{fig:design_study_process}
\end{figure}

Methodologically, we were primarily influenced by the design study methodology~\cite{Sedlmair2012},
but also included elements of classical user-centered design~\cite{Mao2005}, and participatory design~\cite{Boedker2022}.
Our overall goal is to make the faithful communication of drug knowledge more efficient using visualizations to ensure medication safety. In our design study (\Cref{fig:design_study_process}), we aim to investigate what drug instructions are suitable for visualizations and how these visualizations are designed.
Our team comprises members from diverse backgrounds: four members are from the same university, of which three are from the same institute of health data (health data group, hereafter), and one is from the department of journalism and communication---one has 8 years of clinical medicine experience and 3 years of visualization expertise, one has 10 years of experience in pharmacoepidemiology and health data science, one has 15 years of experience in visualization, and one has 18 years of experience in communication. 
Two members are from the other university, of which one has 18 years of experience in human-computer interaction (HCI) and visualization, and the other one has nine years of HCI and accessibility.  
To this end, our interdisciplinary research team collaborated with three pharmacists to conduct design workshops and survey drug databases to develop a framework for visual drug instructions.
The three pharmacists who collaborated with us have 12, 13, and 13 years of pharmaceutical experience, respectively.

\subsection{Decision on Visualization Design for OTC Drugs and a Taxonomy of Drug Instructions}
\label{sec:otc_type}

We explain the rationale of focusing only on visualization design for OTC drugs and then introduce the taxonomy of drug instructions that guide our design.

\subsubsection{Initial Design Attempts}
\label{sec:initial_attempts}
We invited the three pharmacists to select a typical drug for routine clinical use based on the principle of comprehensive content of the drug instructions. 
They chose levofloxacin tablets -- a commonly used prescription drug in clinical practice, and labeled contents needed for different stakeholders. 

\begin{figure}[ht]
  \centering
  \includegraphics[width=\linewidth]{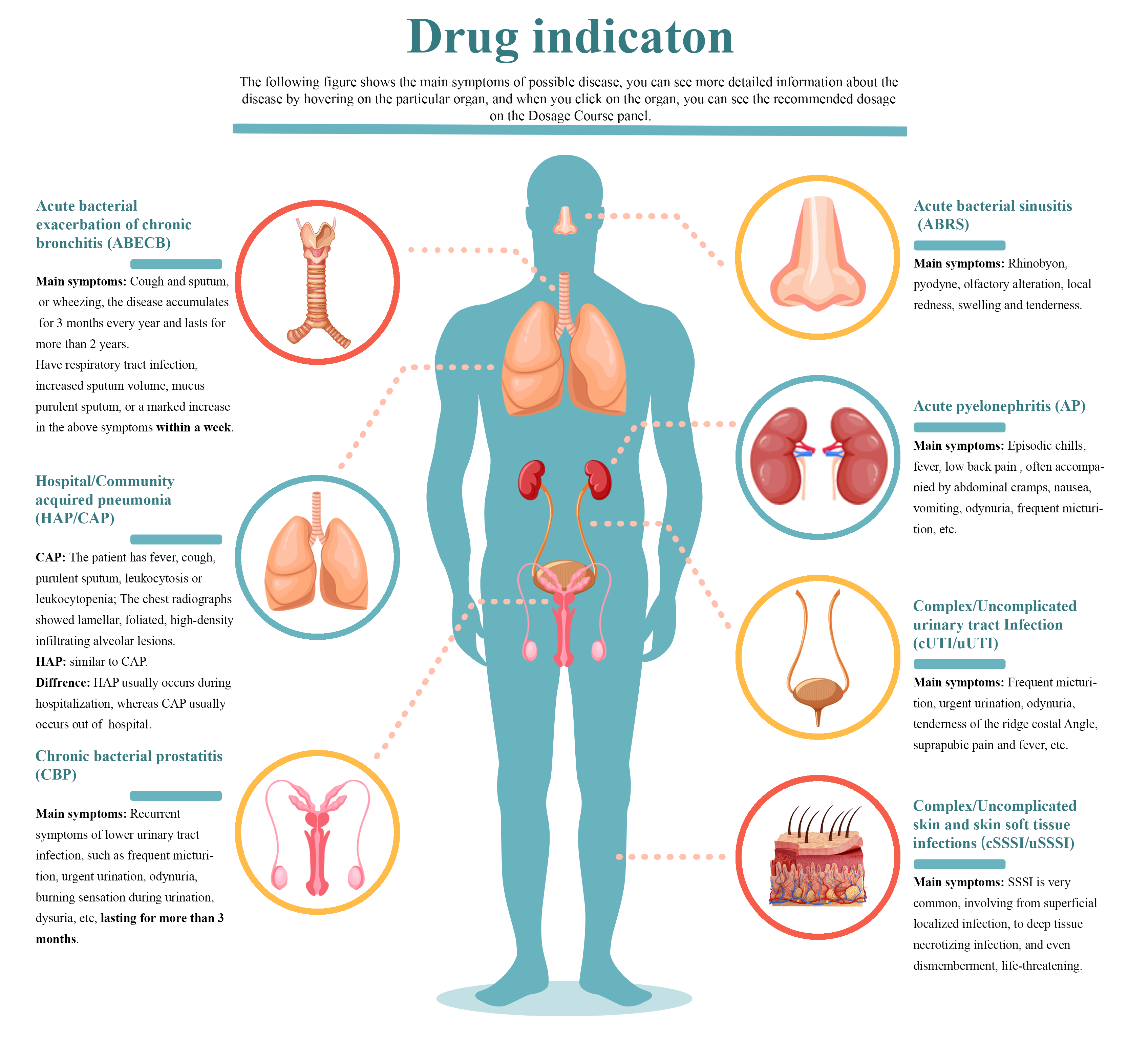}
  \caption{The visualization of levofloxacin's indications poses a comprehension challenge for non-expert users}
  \label{fig:humanOrgans}
\end{figure}

We analyzed the labeled instruction and made a pilot design of visualizations under the guidance of pharmacists for different contents of the instruction. 
A preliminary visualization exploration of levofloxacin is provided in the supplemental material (1.supplemental\_material\_main.pdf). 
After several attempts, we found it challenging to use appropriate visualizations to effectively convey critical information about levofloxacin, including its indications, usage, and dosage. 
This is because the indications primarily involve diseases that are difficult for those without expertise to understand (\Cref{fig:humanOrgans}), and the usage and dosage vary depending on the specific indication.
Discussing with pharmacists, we concluded that the difficulty is due to levofloxacin being a prescription drug that must be diagnosed and prescribed by a professional with extensive knowledge of it before the drug can be administered and used.
This suggests that the value of visualized instructions for prescription drugs may be limited to medical professionals and is not an urgent requirement for all stakeholders.

In contrast, OTC drugs do not require a prescription and can be used by consumers. The instructions serve as the main guidance, indicating that these instructions can be comprehended by those without specialized medical knowledge.
As a result of this discussion on visualization difficulty, especially for indications, we decided that our current visualization design for drug instructions should focus on OTC drugs instead of prescription drugs.

\subsubsection{Taxonomy of Drug Instructions}
\label{sec:otc_type}

\rev{
Our initial design attempts for a prescription drug highlighted the need to systematically understand what types of drug instructions are amenable to visualization. 
We, therefore, developed a taxonomy to categorize drug instructions based on key dimensions that influence visualization design. The taxonomy focuses on the tuple
\[
C = \left(u, i, p\right)\;,
\]
where $u$ denotes ``usage and dosage", $i$ for ``indication'', and $p$ for ``population or age''.
Each element in the tuple $C$ has two levels: $u$---(changes, fixed), $i$---(diseases, symptoms), $p$---(specific, general).
This framework allows us to classify instructions based on how their core information varies, which directly impacts the complexity and approach required for visual design.
}

\rev{
To apply this taxonomy, we systematically investigated OTC chemical drugs in the NMPA~\cite{nmpa_otc2024} database, which contains 733 Class A and 392 Class B drugs (classified by safety level).
}
Since we had to manually read the instructions for classification, we sampled the database with a sampling rate of 10\%, i.e., 73 Class A and 39 Class B drugs were randomly sampled and examined.
Details of the 112 OTC drugs are reported in the supplemental material (2. sampled\_112\_OTC\_drugs.xlsx).

The initial classification of drug instructions was performed by a team member with a clinical medicine background, based on the fundamental structural and informational patterns inherent in drug instructions. 
To ensure inter-rater reliability, this initial classification was independently cross-validated by a second researcher, a health data scientist with expertise in pharmaceuticals. 
Any discrepancies were resolved through discussion until a consensus was reached.

There exists a total of 8 cases by possible combinations. 
However, we grouped all instructions that use diseases as indications into one category, as they involve specific knowledge of diseases, where visualization does not help much for correct comprehension, similar to prescription drugs. 
These include mostly dual-status drugs, for example, Omeprazole Enteric-coated tablets, whose indications include diseases like gastroesophageal reflux disease, which is difficult to convey with visualization.
Then, four cases remain, but the case $(u: \text{fixed}, i: \text{symptoms}, p: \text{specific})$ is equivalent to $(u: \text{fixed}, i: \text{symptoms}, p: \text{general})$, and, therefore, three unique cases exist.

As a result, we classified drug instructions into five elemental categories (Cat.) based on the tuple $C$ (\Cref{fig:design_otc}(1)) and summarization of NMPA database: 
\begin{enumerate}
    \item[Cat. 1] Usage and dosage vary with indications (mainly symptoms).
    \item[Cat. 2] Usage and dosage vary with age/population. 
    \item[Cat. 3] Usage and dosage are fixed.
    \item[Cat. 4] The indications include diseases.
    \item[Cat. 5] Prescription drugs.
\end{enumerate}
We focus on the visualization of OTC drug instructions whose indications are symptoms (Cat. 1-3).
Note that some drugs may be classified into more than one category, for example, the usage and dosage do not vary for different populations for one indication, but vary based on population for another indication (Cat. 1 + 2).

\subsection{Visualization Design Requirements for OTC Drug Instructions}
\label{sec:design_req}

\rev{
To translate the high-level focus on OTC drugs into concrete design requirements, we conducted a structured participatory design workshop with the three collaborating pharmacists, using paracetamol sustained-release tablets as a case study. 
The process was guided by a tailored medical visualization model~\cite{Fan2025} and facilitated with Five Design Sheets (FDS)~\cite{Roberts2016} for ideation and VizItCards~\cite{He2017} for rapid sketching and critique.
}

\rev{
\textbf{Workshop activities and links to requirements:} The workshop began with problem discussions, leading to brainstorming and the creation of low-fidelity sketches exploring layouts, icons, and methods for linking information. 
These sketches were discussed, grouped, and evaluated through card-sorting exercises. 
We also conducted structured prioritization activities, such as Likert-scale scoring of section importance. 
The following four requirements (R1–R4) were consolidated directly from these tangible outputs---specific sketches, card-sort groupings, and priority lists---and served as the direct justification for our subsequent visualization design choices.
}

    \paragraph{\textbf{R1:} Design drug instructions for different audience}
    \rev{
    \textbf{Requirement:} Provide two information views: a simplified version for the public (e.g., patients) and a complete version for medical professionals (e.g., pharmacists).
    \textbf{Causal chain from the workshop:} This foundational need emerged from initial discussions and was reflected in multiple sketches, which consistently showed two layouts or a toggle mechanism, rejecting a ``one-size-fits-all'' approach. 
    A collaborative card-sorting exercise specifically defined the ``must-have'' content for the simplified version (drug name, characteristics, specifications, indications, usage and dosage, contraindications, precautions, storage, and expiry date) and ``expert-only'' information for the complete version. 
    This output directly justified implementing two distinct interface modes to serve different health literacy levels~\cite{health-literacy2025}.
    }
    
    \paragraph{\textbf{R2:} Arrange the layout according to the priority of sections}
    \rev{
    \textbf{Requirement:} The visual layout must spatially reflect the relative priority of information sections.
    \textbf{Causal chain from the workshop:} Early sketches highlighted the challenge of allocating screen space. 
    A structured prioritization activity resolved this: participants scored instruction sections. 
    ``Indications'' and ``Dosage and Usage'' received the highest average scores (>4.0), followed by ``Precautions'' with pharmacists uniquely emphasizing ``Drug Interactions''. 
    This quantitative output became the non-negotiable constraint for our layout, mandating a top-down structure that places the most critical content most prominently, consistent with user-centered design principles~\cite{Mao2005} and prior research~\cite{Raynor2007}.
    }
    
    \paragraph{\textbf{R3:} Correspond the usage and dosage to indications and populations}
    \rev{
    \textbf{Requirement:} The visualization must explicitly correlate specific ``Usage and Dosage'' information with the corresponding ``Indication'' and/or ``Patient Population.''
    \textbf{Causal chain from the workshop:} Analysis of early sketches for paracetamol revealed confusion in depicting conditional dosing. 
    Pharmacists identified this as a high-risk area for misunderstanding in text. 
    A dedicated brainstorming session using VizItCards then focused on ``linkage'' mechanisms, with concepts like color-coding and interactive selection being sketched and evaluated. 
    The consensus favored a clear, interactive visual link over static text, directly leading to the design of interactive linkages (e.g., click indication → update dosage) in the prototype to mitigate safety risks.
    }
    
    \paragraph{\textbf{R4:} Design appropriate visualization for each section}
    \rev{
    \textbf{Requirement:} The visualization technique for each section must be appropriately matched to the type of information it conveys (e.g., icons for warnings, hierarchies for interactions).
    \textbf{Causal chain from the workshop:} During sketching, participants intuitively proposed different graphical forms for different content. 
    A focused critique session evaluated these against goals like reducing cognitive load and improving understanding. 
    The workshop produced a specific mapping: categorical warnings(e.g., Precautions) → standardized pictograms; hierarchical information(e.g., Drug Interactions) → tree-based visualizations; numerical dosage→ clear typography with pictographic reinforcement. 
    This mapping directly specified the visual encoding used for each section in the final design, informed by information design best practices~\cite{Rezaie2024}.
    }

\subsection{Initial Prototype of an OTC Drug Instruction}
\label{sec: prototype}

\rev{
To translate the design requirements (Section~\ref{sec:design_req}) into a concrete, testable artifact, we developed an initial interactive visualization prototype for the drug instruction of paracetamol sustained-release tablets. 
A conceptual overview of the complete version is provided in \Cref{fig:initial_prototype}  (see also the larger version in the supplementary material), illustrating the key design features and layout scheme rather than the complete informational content.
}

\begin{figure}[ht]
  \centering
  \includegraphics[width=\linewidth]{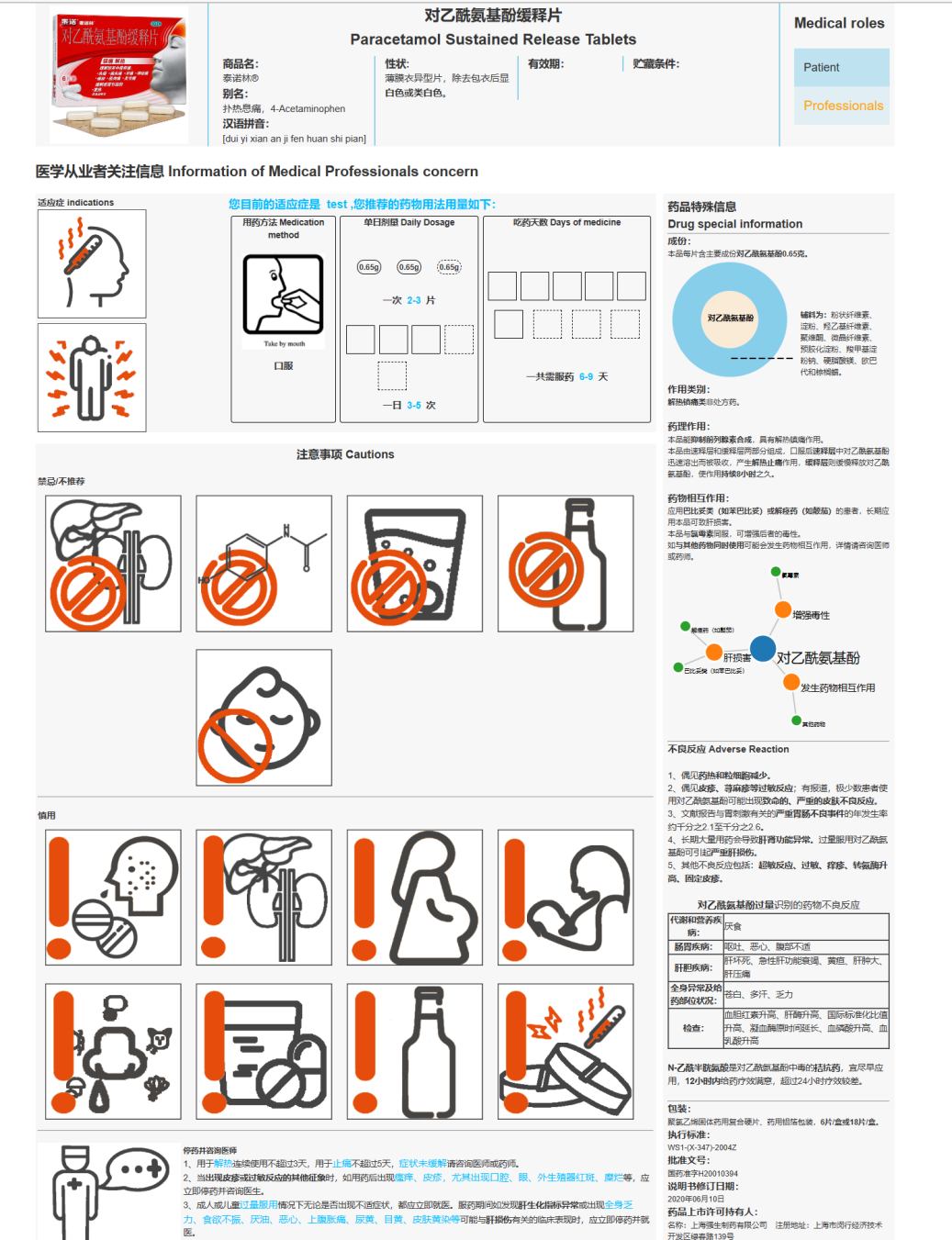}
  \caption{A conceptual overview of the initial prototype (complete version) for paracetamol in Chinese.}
    \label{fig:initial_prototype}
\end{figure}

\subsubsection{Indications, Usage and Dosage}
\rev{
\textbf{Design problem and rationale:} This section implements requirements R2 (priority layout) and R3 (dosage-indication linkage). 
The core challenge was to present conditional dosing (different doses for different symptoms) clearly and compactly. 
Based on requirement R3, we rejected a static, text-heavy list. 
Instead, we implemented an interactive selection mechanism: clicking a symptom pictogram updates the dosage display. 
This creates a direct, unambiguous link, addressing the safety risk of mismatched information identified in the workshop.
} 

\rev{
\textbf{Visual encoding of dosage frequency:} To represent dosage frequency (e.g., once daily, multiple times per day) (\Cref{fig:initial_prototype}) or duration, we used a sequence of rectangular units. A key design decision was how to represent ranges like ``3--5 times per day.''
\begin{itemize}
    \item \textbf{Alternatives considered:} We considered simple text, a numeric slider, or a graduated bar. Text was deemed less scannable; interactive widgets were overly complex for core information display.
    \item \textbf{Chosen design \& rationale:} We used a combination of solid and dashed rectangles (e.g., three solid + two dashed). 
    This encoding leverages a common metaphor (solid=confirmed, dashed=optional/possible) to compactly convey flexibility. 
    It aligns with requirement R4 by matching a visual metaphor to a specific data property (uncertain range). 
    We assume users have basic visual literacy to interpret this common metaphor. 
    The primary benefit over text (``3–5 times per day'') is immediate visual segmentation of the range, potentially reducing the cognitive step of parsing the numeric span.
\end{itemize}
}

\subsubsection{Precautions}
\rev{To address the complexity and potential confusion within the standard ``Precautions'' section, we focused on categorizing and visually distinguishing warnings by severity.} 
We found that this drug instruction has adverse reactions, precautions, and contraindications, all within the precautions section.
This can be confusing to users, potentially leading them to overlook or skip important drug safety information.
Therefore, it is necessary to communicate this information correctly and clearly to users, especially to specific populations such as the elderly and pregnant women.

\begin{figure}[htb]
  \centering
\subfloat[]{\includegraphics[width=0.25\linewidth]{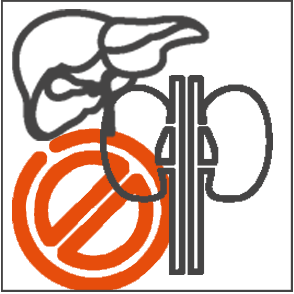}}\label{fig:pre_subfigs_taboo}\hspace{0.5em}
\subfloat[]{\includegraphics[width=0.25\linewidth]{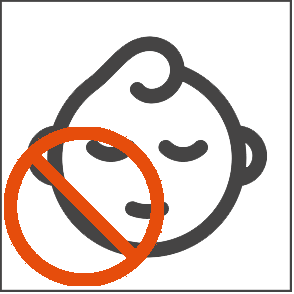}}	\label{fig:pre_subfigs_norec}\hspace{0.5em}
\subfloat[]{\includegraphics[width=0.25\linewidth]{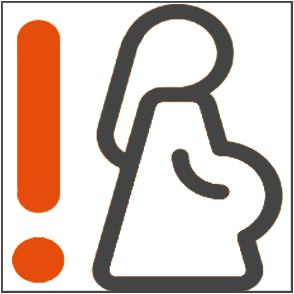}}\label{fig:pre_subfigs_caution}\\
\subfloat[]{\includegraphics[width=0.25\linewidth]{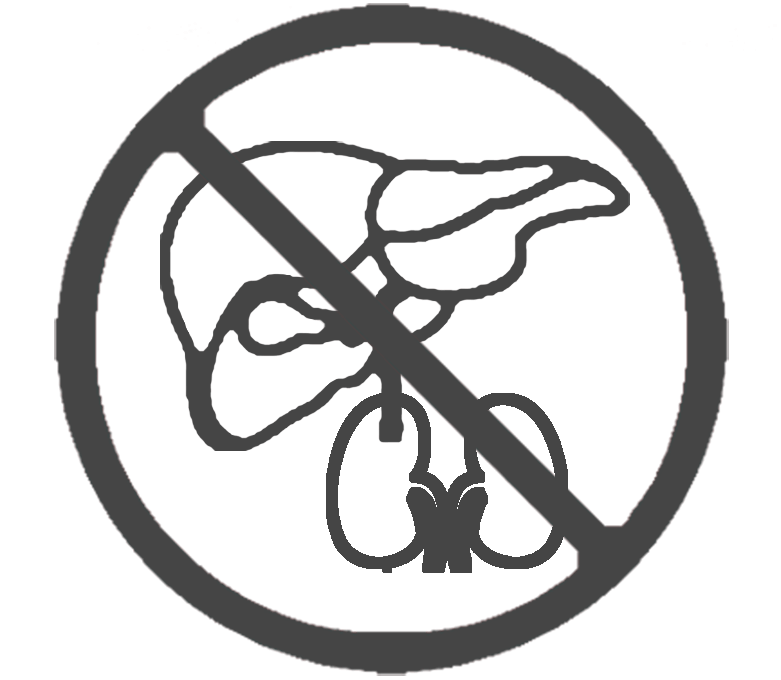}}\label{fig:subfigs_taboo}\hspace{0.5em}
\subfloat[]{\includegraphics[width=0.25\linewidth]{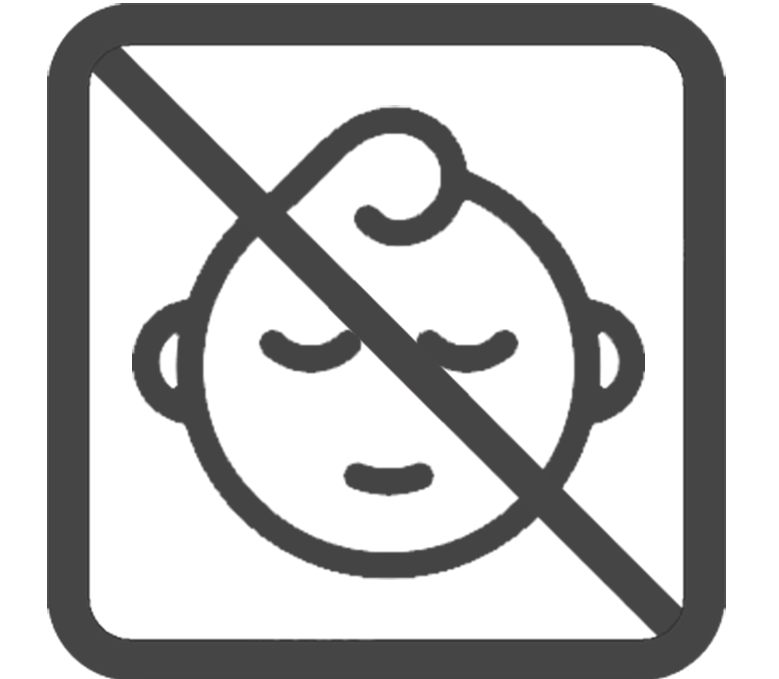}}\label{fig:subfigs_norec}\hspace{0.5em}
\subfloat[]{\includegraphics[width=0.25\linewidth]{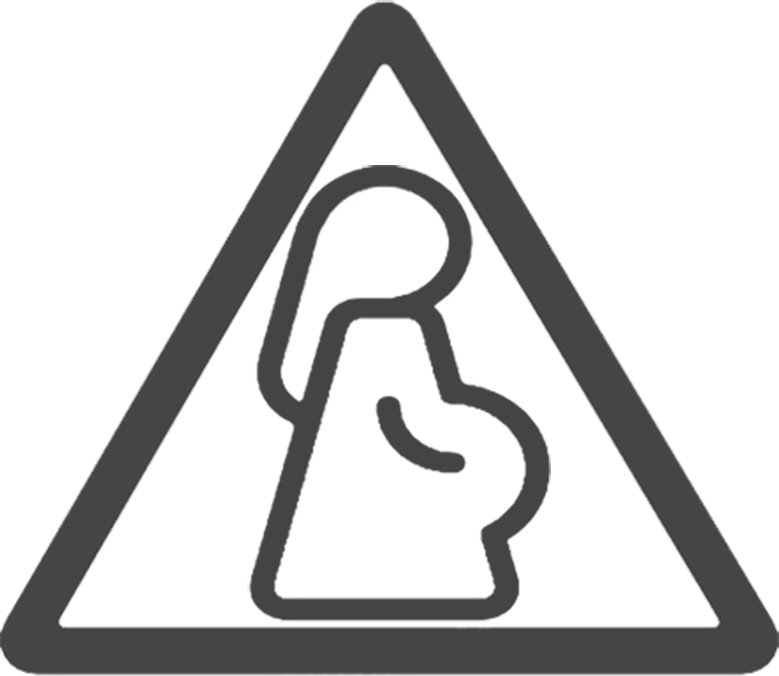}}\label{fig:subfigs_caution}
  \caption{Examples of three categories of precautions---contraindication (a,d), not recommended/allowed (b,e), and use with caution (c,f)---in the initial (the top row) and revised (the bottom row) prototype.}
  \label{fig:subfigs_precautions}
\end{figure}

After consulting with the pharmacists, we classified precautions into three categories by the severity of the warning: \emph{contraindications}, \emph{not recommended/not allowed}, and \emph{use with caution}.
As shown in~\Cref{fig:subfigs_precautions} (top row), we designed different legends for these three categories to help with better understanding, where a double circle with a slash represents \emph{contraindications}, an exclamation mark represents \emph{use with caution} and a single circle with a slash represent a less clear intermediate state, defined as \emph{not recommended/not allowed} in our design. 
In addition, some drugs may need to be stopped under certain circumstances. 
Therefore, except for the three categories, we also summarized the situation of ``discontinue the drug and consult a physician.'' 
In the initial prototype, we designed a ``doctor with speech bubble'' pictogram (the lower left corner of \Cref{fig:initial_prototype}) to highlight this situation, with the relevant content emphasized only in key sections using color highlighting.

\subsubsection{Drug Interactions}
\rev{
\textbf{Design problem and rationale:} This section responds to the pharmacists' emphasis (R2) and the need for appropriate visual forms (R4). 
Discussions revealed that the drug interaction information in current medication instructions suffers from an ambiguous structure and weak logical connections, which increases the cognitive burden of users.
To address this, we adopted a hierarchical organizational framework consisting of three top-level categories: effects of other drugs on the current drug, effects of the current drug on other drugs, and mutual drug interactions. 
Specific details---such as mechanisms, severity, and clinical advice---are organized as child nodes within this hierarchy.
} 

\begin{figure}[!htb]
  \centering
  \subfloat[Icicle Plot]{%
    \includegraphics[width=0.64\linewidth]{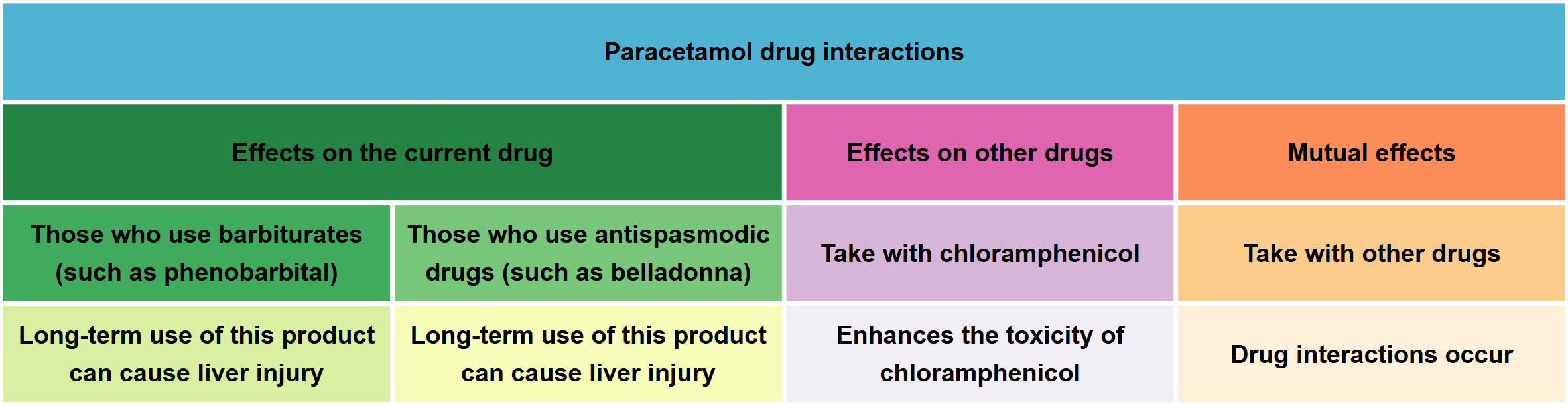}%
    \label{fig:icicle_plot}%
  }
  \subfloat[Force-Directed Tree]{%
    \includegraphics[width=0.34\linewidth]{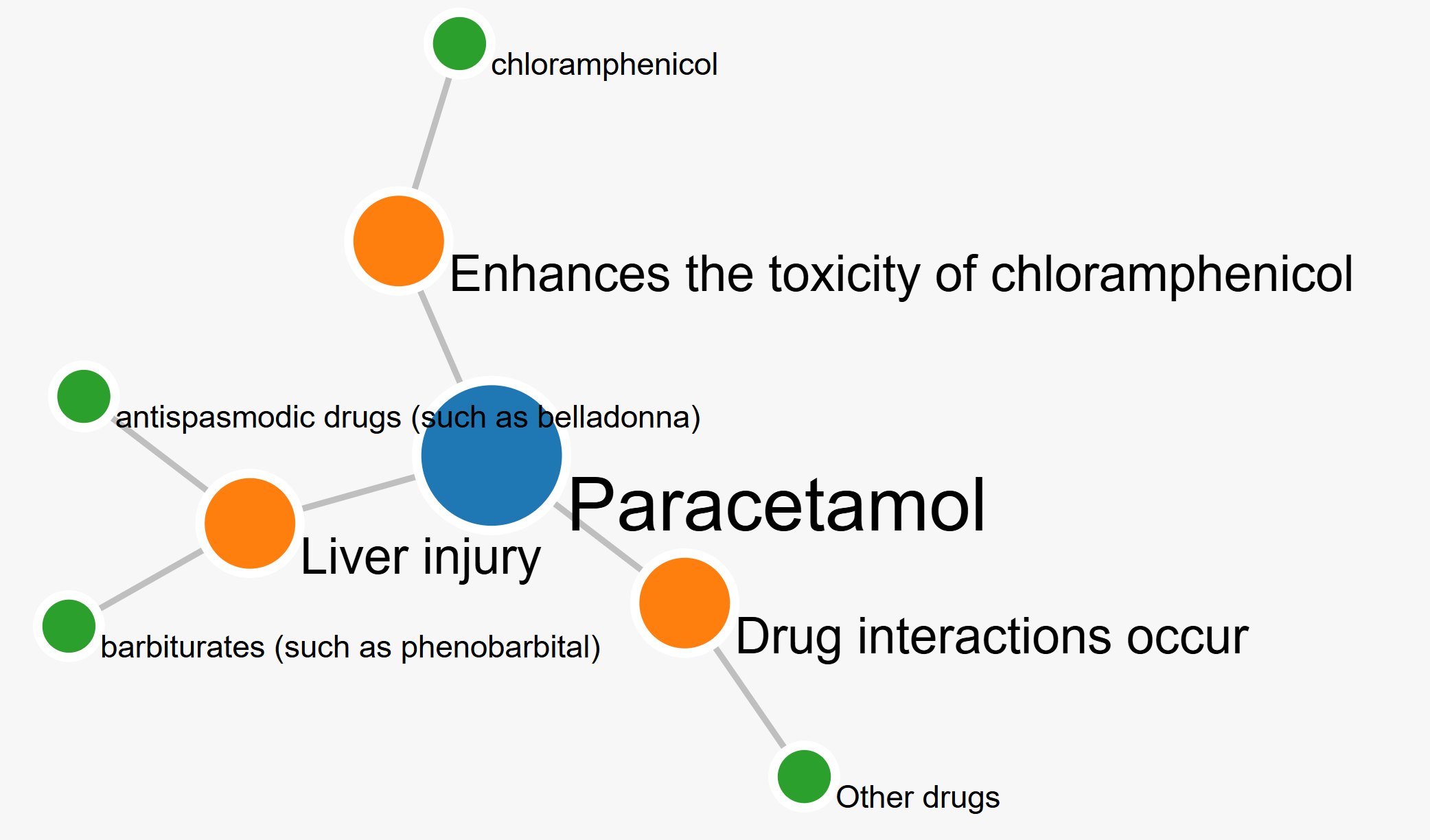}%
    \label{fig:force_directed_tree}%
  }\\
  \subfloat[Cluster Tree]{%
    \includegraphics[width=0.64\linewidth]{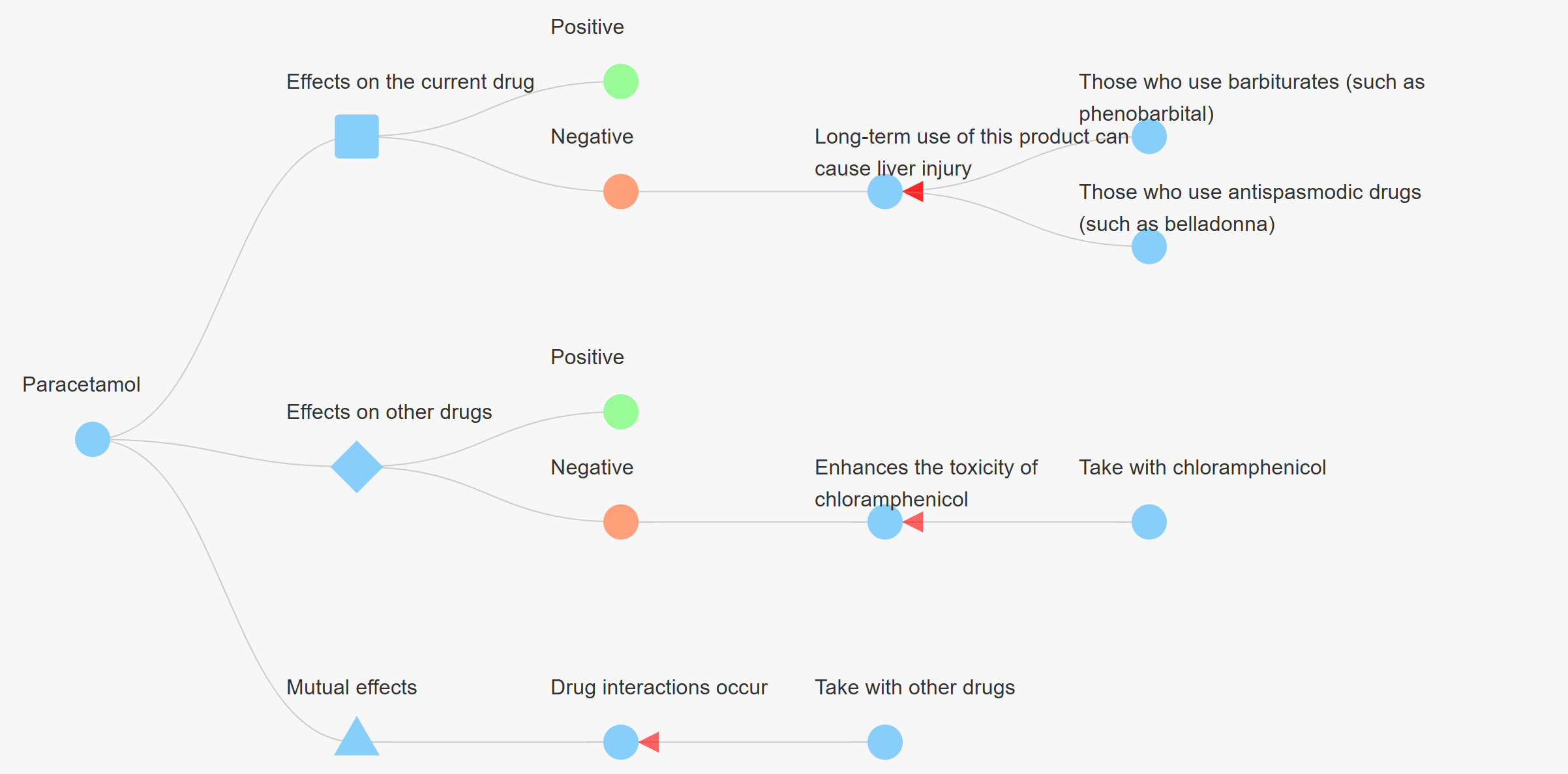}%
    \label{fig:cluster_tree}%
  }
  \subfloat[Circle Packing]{%
    \includegraphics[width=0.34\linewidth]{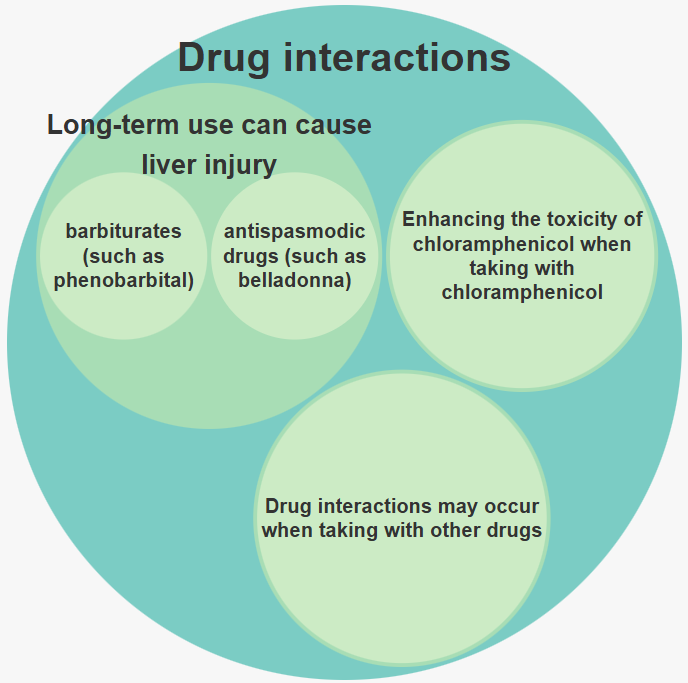}%
    \label{fig:circle_packing}%
  }
  \caption{Four complementary visualization forms for drug interactions of paracetamol.}
  \label{fig:four_interaction_type}
\end{figure}

\rev{
Guided by requirement R4 (match form to content), we needed a visualization suited for hierarchical data.
\begin{itemize}
    \item \textbf{Explored alternatives:} We internally sketched and gathered preliminary pharmacist feedback on four candidate techniques:
    An icicle plot provides an intuitive top-down view of hierarchical flow, supporting path tracing across levels (see \Cref{fig:icicle_plot}).
    A force-directed tree visually organizes nodes to highlight central ones and connection strengths, helping to identify key interactions (see \Cref{fig:force_directed_tree}).
    A cluster tree highlights categorical grouping and subordination, clarifying taxonomic relationships (see \Cref{fig:cluster_tree}).
    Circle packing employs a space-filling layout to compactly encode the full hierarchy (see \Cref{fig:circle_packing}).
    \item \textbf{Chosen design \& rationale:} The icicle plot was provisionally selected for deeper development. It provides a clear, space-filling representation of the hierarchy, making the depth and breadth of interaction categories immediately apparent. This aligns with R4 by applying a hierarchical visualization to hierarchical data. Its relative simplicity supports our assumption of moderate visual literacy among users seeking detailed drug information. Interactive tooltips provide the exact textual information on hover, ensuring accuracy is maintained.
    The colors in the icicle plot were primarily used to visually distinguish between different hierarchical levels of the drug interaction information. In our design, each distinct color represented a major category or branch at a specific level of the taxonomy. Within a branch, a consistent hue with varying saturation or lightness may be applied to indicate sub-categories, thereby maintaining visual grouping while showing depth.
\end{itemize}
} 

\rev{
Guided by the ``overview first, details on demand''~\cite{Shneiderman1997} principle, the icicle plot, supplemented with icons and interactive details, aims to facilitate both high-level understanding and detailed exploration.}

\subsection{Feedback on Initial Prototype}
\label{sec: feedback}
\rev{To validate and refine the initial prototype, we conducted a feedback session with the same three pharmacists. The primary goal was to gather concrete critiques and suggestions to guide the next iteration of our visualization design. The feedback was comprehensive, covering visual encoding, layout, and content accuracy, and directly informed the revisions described in the following section.}

Regarding legends and pictograms, participants suggested expanding the information conveyed within rectangles to enhance clarity on medication methods, frequency, and duration. 
They also recommended redesigning precaution legends, drawing inspiration from traffic signs for better intuitiveness.
Additionally, it was proposed to use pre-attentive visual elements to help users quickly identify changes---such as in medication duration---when switching between different drug indications.
Furthermore, for the ``discontinue the drug and consult a physician'' scenario, participants suggested adopting a similar icicle-based visualization, as this scenario also involves categorical hierarchies akin to drug interactions, thereby ensuring consistency in design logic and user comprehension.

For layout, it was agreed that professional information should be placed below the core contents (indications, usage, dosage, and precautions), as different user groups share the same primary concerns, despite having varied secondary needs.
Other feedback included avoiding the use of drug-specific packaging imagery due to copyright concerns; renaming the navigation labels from ``patient'' and ``pharmacists'' to ``simplified'' and ``complete'' to better reflect the two instruction formats; removing the double-ring chart for ingredient information, as it occupies considerable space without significantly enhancing comprehension; removing the module for medication methods, as it does not change with the drug; and addressing contradictions such as ``once every 8 hours'' versus ``three times a day''. Pharmacists emphasized that all information in drug instructions must remain legally accurate without oversimplification.

\rev{This feedback provided a clear plan for improving the prototype, making the next design iteration more user-friendly, legally compliant, and focused on user needs.}

\subsection{Revised Prototype}
\begin{figure*}[!tb]
\captionsetup[subfigure]{labelformat=empty}
  \centering
  \subfloat[Simplified]{\includegraphics[height=14cm]{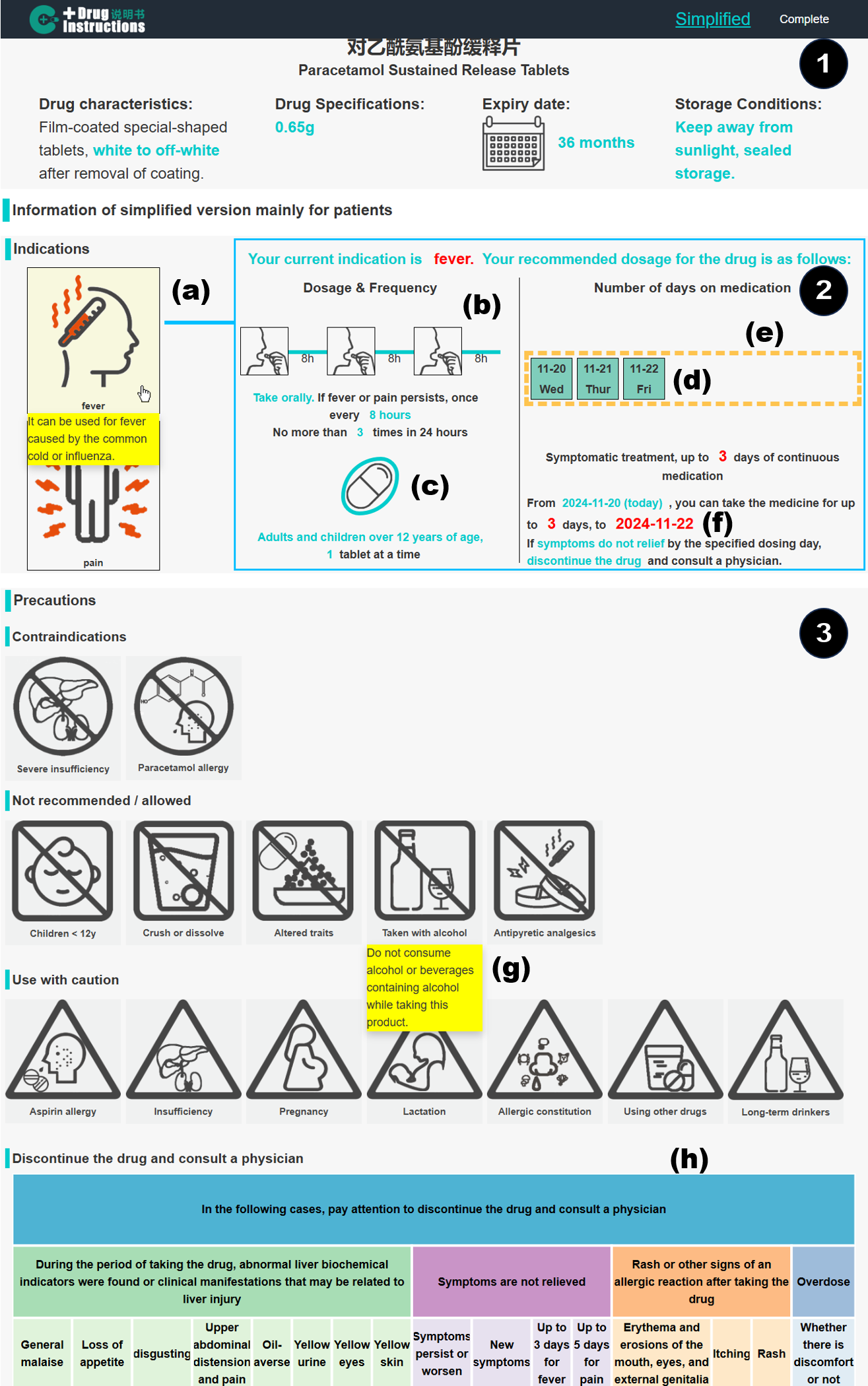}}\hspace{0.3em}
  \subfloat[Complete]{\includegraphics[height=14cm]{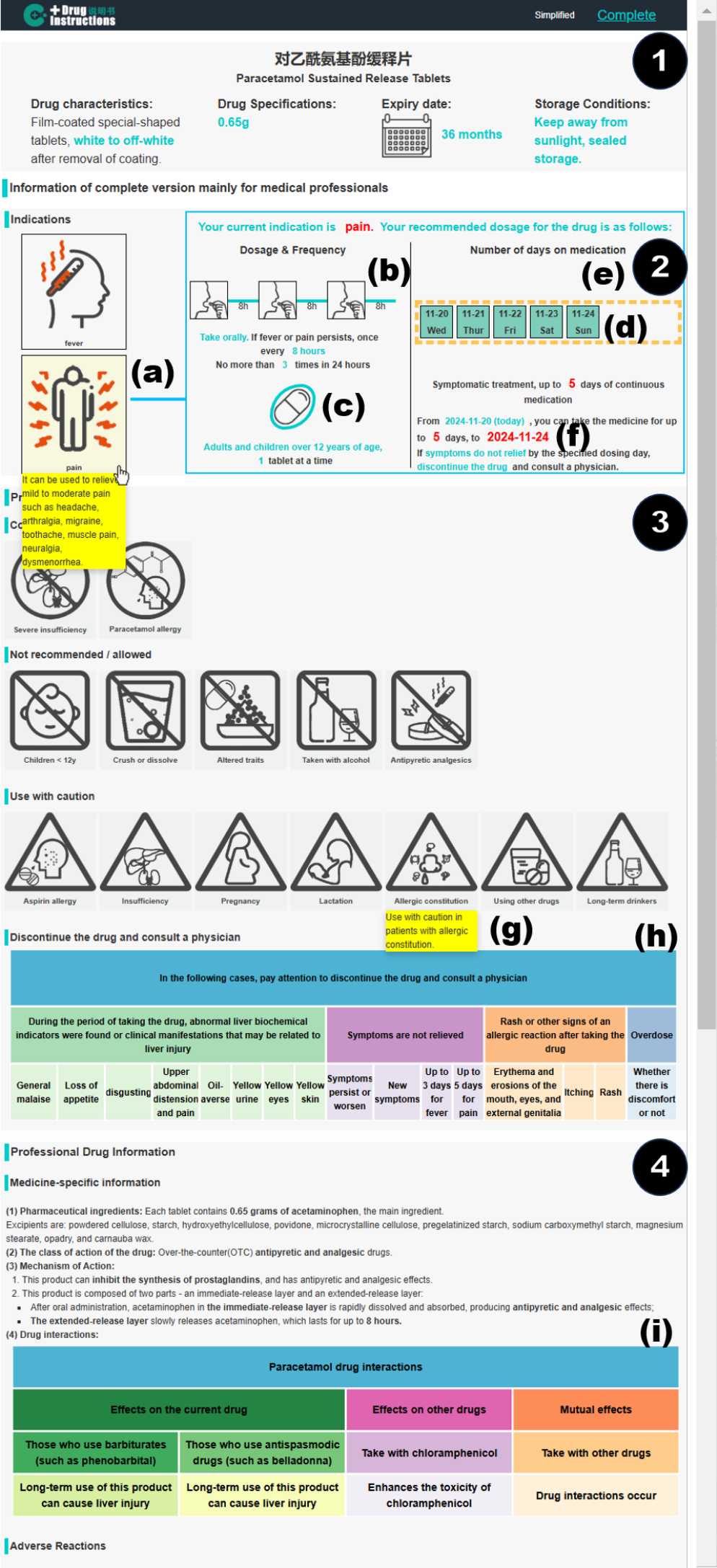}}
  \caption{The (left) simplified and (right) complete versions of the visualization design for paracetamol. Three common modules are included in both versions: (1) The \emph{basic information and navigation module} allows the user to switch between the simplified and complete versions. (2) The \emph{indication and usage module} allows the user to click on a specific indication to show the corresponding dosage and usage. (3) The \emph{precautions module} shows different precautions by pictograms and an icicle plot (h). Specifically for the complete version, (4) the \emph{professional drug information module} includes an icicle plot representation of drug interactions (i). 
       }
  \label{fig:revised_design}
\end{figure*}

We modified our prototype based on the feedback, and the simplified and complete versions of the revised visualization are shown in~\Cref{fig:revised_design}.
[We will provide the link to the interactive visualization tool upon acceptance.]

In our revised visualization, icons for medication method (take orally,~\Cref{fig:revised_design}(b)), drug characteristics (tablet,~\Cref{fig:revised_design}(c)), frequency (\Cref{fig:revised_design}(b)) and days (\Cref{fig:revised_design}(d) showing the date of first taking a specific medication and the duration of days) are included to help understand the usage and dosage.

Inspired by pharmaceutical pictograms of USP and the design of traffic signs, we revised our three-category precautions legends (\Cref{fig:subfigs_precautions}(bottom row), \Cref{fig:revised_design}(3)), which use a circle with a slash representing \emph{contraindication}, a triangle indicating \emph{use with caution}, and a rectangle with a slash representing \emph{not recommended/allowed}.
Overlaying specific categories of precaution legends with their corresponding precaution content resulted in our final visualized version of the precautions.
The principle of pre-attention was applied to highlight the changed information when switching between indications to help users identify differences quickly (changed contents in red font and dotted orange boxes as in~\Cref{fig:revised_design}(e, f)). This principle can be applied to any changed content for highlighting.
For simplified information shown in the pictograms, the full, unmodified contents from drug instructions can be displayed when hovered over or clicked (\Cref{fig:revised_design}(a, g)).

\begin{figure}[!tb]
  \centering
  \subfloat[Usage and dosage vary with indications--the example of paracetamol. When users click on one of the indications, the visualization updates accordingly.\label{subfig:dosage_indication}]{\includegraphics[width=\linewidth]{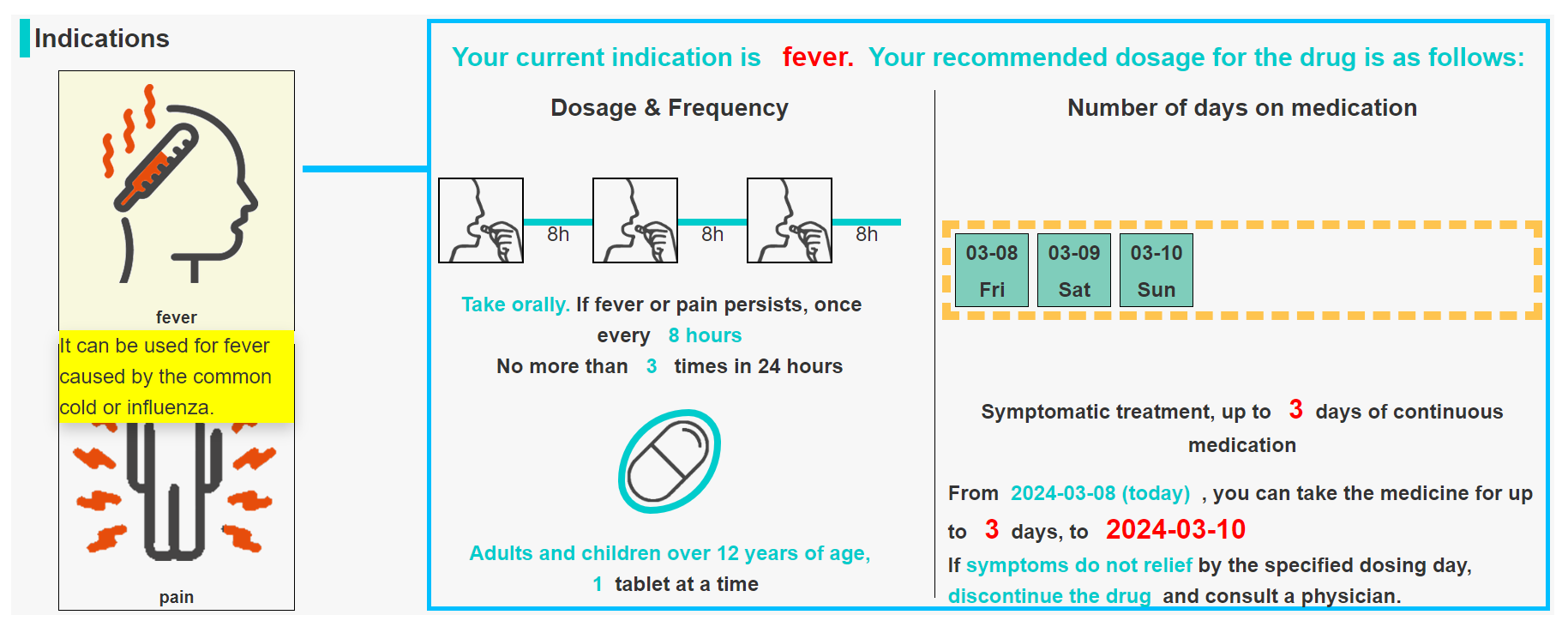}}\hfill
  \subfloat[Usage and dosage vary with age/population--the example of cetirizine. The usage is highlighted (orange) accordingly when the user hovers over it.\label{subfig:dosage_population}]{\includegraphics[width=\linewidth]{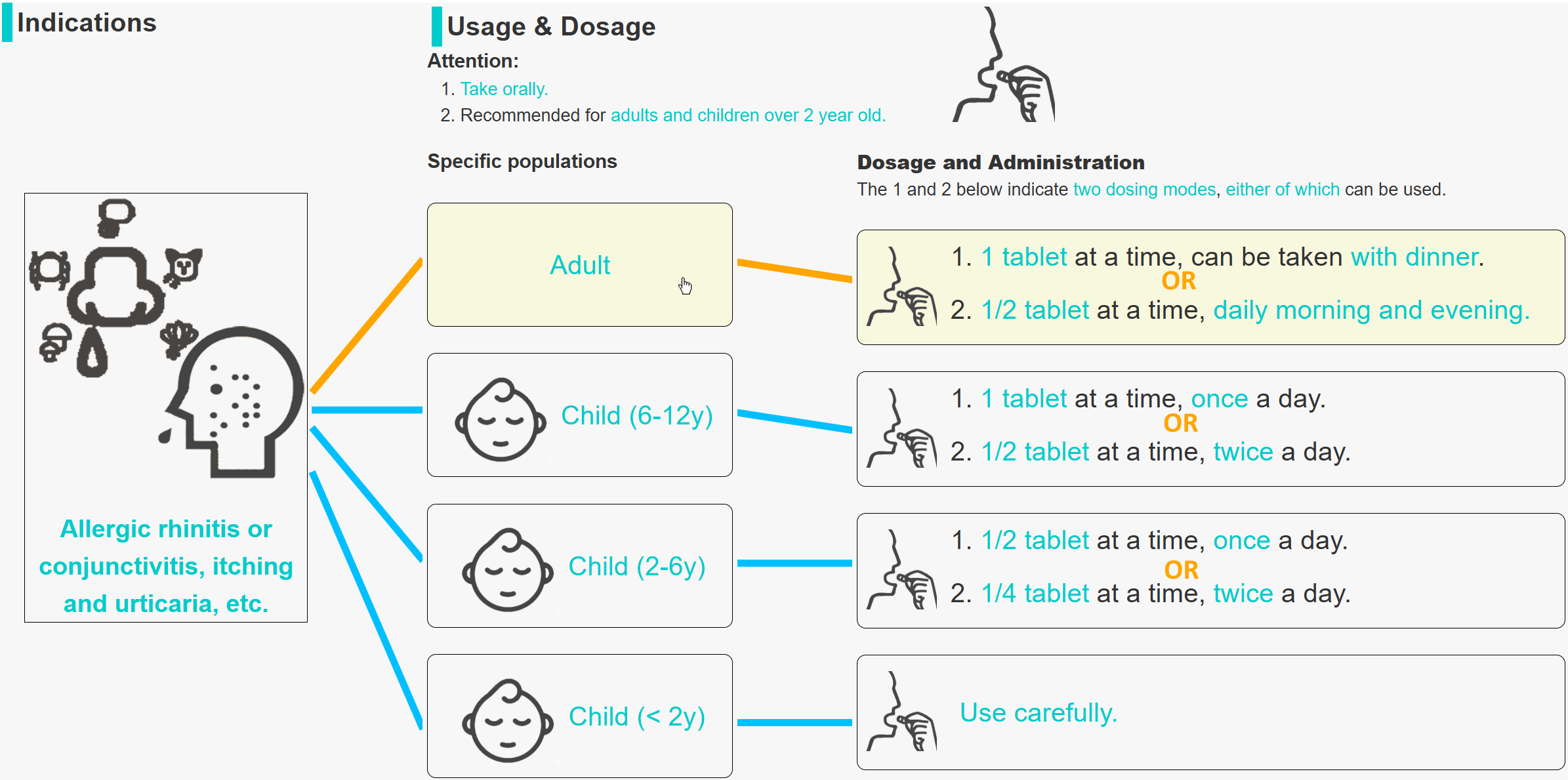}}\\
  \subfloat[Usage and dosage fixed--the example of Ambroxol.\label{subfig:dosage_fixed}]{\includegraphics[width=\linewidth]{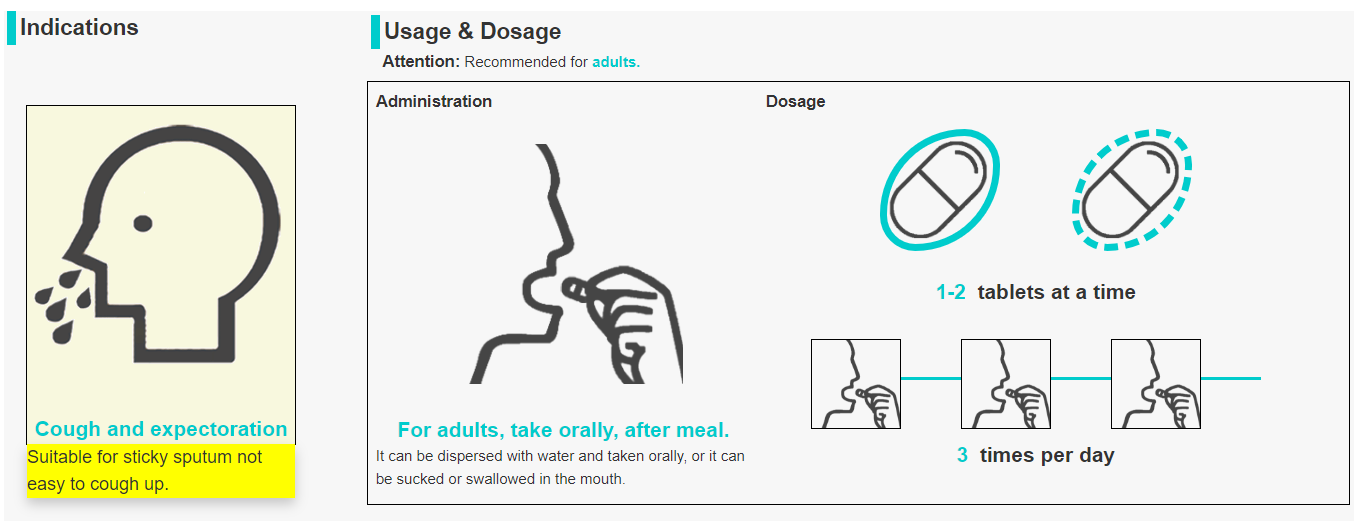}}  
  \caption{Visualization designs for indication, usage, and dosage of three categories: (a) Cat. 1, (b) Cat. 2, and (c) Cat. 3 of the OTC drugs.}
  \label{fig:three_dosage_type}
\end{figure}

The difference between the two versions lies in the fact that the simplified version only contains modules (1)---(3), without module (4), which only shows in the complete version.
Users can switch between different versions through the navigation button in the upper-right corner.

\subsection{Designs for Other OTC Categories}
\rev{Our taxonomy (\Cref{sec:otc_type}) defined three OTC categories (Cat. 1-3) suitable for visualization. 
Having developed an initial prototype for Cat. 1 (paracetamol), we needed to validate and adapt our design framework for the remaining categories to ensure its generalizability. 
This subsection describes the application of our framework to representative drugs from Categories 2 and 3.}

With input from the pharmacists, we selected two widely used OTC drugs as representative cases: 
\begin{itemize}
\item \textbf{Cetirizine dihydrochloride tablets (Cat. 2):} Characterized by dosage that varies with age/population.
\item \textbf{Ambroxol hydrochloride dispersible tablets (Cat. 3):} Characterized by a fixed dosing regimen.
\end{itemize}
These drugs were chosen because they are clinically common, have clearly differentiated dosing patterns characteristic of their categories, and serve as ideal test cases for evaluating our visualization framework.

The core design adaptation across categories lies in how the ``Usage and Dosage'' section visually links to indications or populations (\Cref{fig:revised_design}(2))\rev{, directly implementing requirement R3}.
Other contents, such as precautions and drug interactions, follow consistent design patterns, with only specific content varying by drug.

\rev{
\begin{itemize}
    \item \textbf{Cat. 1 (e.g., Paracetamol):} Uses an \textbf{interactive selection} mechanism. Clicking on a specific indication updates the corresponding dosage display (\Cref{subfig:dosage_indication}), managing space for multiple indication-dosage pairs.
    \item \textbf{Cat. 2 (e.g., Cetirizine):} Uses a \textbf{decision tree} to visually guide the user from indication, through specific population (e.g., age group), to the corresponding dosage (\Cref{subfig:dosage_population}).
    \item \textbf{Cat. 3 (e.g., Ambroxol):} Presents the indication and fixed dosage in a simple \textbf{side-by-side} layout (\Cref{subfig:dosage_fixed}).
\end{itemize}
This structured approach demonstrates how our framework produces distinct yet coherent visualizations tailored to the logical relationships inherent in each OTC category.}
It should be noted that some drug instructions may require consideration of both Cat. 1 and Cat. 2 during visualization design.

\subsection{Expert User Evaluation on the Revised Prototype and the Taxonomy of OTC Drug Instructions}
\label{sec:expert_evaluation}

\rev{
To assess the clinical validity and appropriateness of the revised prototypes and the proposed taxonomy, we conducted a summative evaluation session with our three collaborating pharmacists. 
The goal was to gather expert feedback on whether the design framework and its category-specific instantiations were coherent, accurate, and aligned with professional practice.
}

\rev{\textbf{Method:}} Using the think-aloud method while presenting the taxonomy and the three interactive visualization interfaces (corresponding to the design variants in~\Cref{fig:three_dosage_type}(a–c)), we collected and documented their feedback on both the conceptual framework and the concrete designs.

\rev{
\textbf{Expert feedback and results.} The pharmacists' feedback focused on their domain of expertise---clinical utility and information accuracy---and is summarized as follows.
\begin{itemize}
    \item \textbf{Taxonomy \& framework validity:} The pharmacists confirmed that the taxonomy logically categorizes OTC instructions based on key variables (dosage, indication, population) that align with their clinical reasoning. They agreed that the three categories (1-3) cover the common, visualizable OTC scenarios and that the proposed design strategies for each are appropriate.
    \item \textbf{Design effectiveness \& clinical usefulness:} They responded positively to the visual designs, stating that each variant correctly implemented the logic of its category (interactive selection, decision tree, side-by-side). They specifically endorsed the classified precaution icons for clarifying warning severity and the hierarchical icicle plot for structuring drug interaction information, noting these features could help prevent common misunderstandings. They judged the visual presentations to be clearer and more scannable than traditional text-based instructions.
\end{itemize}
}

\rev{
This expert validation confirmed the clinical soundness of our taxonomy and framework. 
Note that the decision to evaluate the core usability of the shared framework components through a single, representative drug (paracetamol, Category 1) in the subsequent user study was a methodological choice by the research team.
It was based on the observation---confirmed by the feedback of pharmacists---that the core layout, interaction principles, and visual components for all sections except the dosage-indication linkage were consistent across categories.
}

\subsection{Implementation}
Our work was designed to be used in a web browser for easy accessibility.
The visualizations and the user interface were implemented in Javascript aided by the d3.js~\cite{d32024}
visualization library.
We referred to icon libraries like FLATICON~\cite{FLATICON2024},
Font Awesome~\cite{fontawesome2024},
The Noun Project~\cite{nounproject2024}
and applied Adobe Photoshop to design and modify the various pictograms/icons used in this study under the guidance of pharmacists.

\section{Evaluation of the Visualization Design}
\rev{To objectively evaluate the effectiveness of our proposed visualization framework, we conducted a controlled user study.}
Given that the three design variants differ primarily in the dosage‑indication relationship (the linkage mechanism) while sharing the same layout, interaction principles, and visual components for other instruction sections, we selected the most common and clinically representative case---paracetamol (Category 1)---for the controlled user study. 
This allowed us to evaluate the core visualization framework in depth while ensuring a focused and feasible experimental design. 
The expert evaluation described above (see~\Cref{sec:expert_evaluation}), which confirmed the clinical appropriateness of all three variants, supports the representativeness of this single-case evaluation for the shared framework components.

The study was approved by the biomedical ethics committee of the University of the first author.
\rev{
Before the formal study, we conducted two rounds of pilot studies with three volunteers in each group for a total of 12 participants.
All pilot participants were students or postdoctoral researchers at the first author's university. 
Key changes based on feedback included: enlarged the toggle button for switching between instruction versions to improve discoverability; split the knowledge test into ten blocks with a break to reduce fatigue.
These refinements improved the overall validity and participant experience of the formal study.
}

\subsection{Study Design}
A between-subject design ($N$=60) was used for the study.
The main factor of the study was the representation method of the instruction: with the traditional text version (\emph{Text}, $N$=30) and our complete visualization version (\emph{Vis}, $N$=30) being the two levels. 
The performance of participants was tested via a questionnaire consisting of knowledge questions about paracetamol, subjective questions on usability and cognitive load, and subjective comments on the current tested version.

Our hypotheses were as follows. Compared to the text group, the visualization group would have:
\begin{enumerate}[label=\textbf{H\arabic*}]  
    \item \textbf{Higher score improvement in the knowledge questions.} Participants can understand medical knowledge more easily with visualization.
    \item \textbf{Shorter response time for the knowledge questions.} With visualization, participants can read instructions more efficiently and spend less time.
    \item \textbf{Higher usability rating for the drug instruction}. The user interface of the visualization group is more intuitive, easier to understand, and has the potential to support intuitive decision-making and analysis.
    \item \textbf{Higher learnability rating for the drug instruction.} Visualization potentially conveys complex data more easily.
    \item \textbf{Lower cognitive load.} Visualization has the potential to reduce information overload. 
\end{enumerate}

The secondary factor of the study was the different visualization versions of the instruction:  two levels of the complete version and the simplified version, which were tested using the same 30 people.
The applicability of these two versions was tested via subjective questions on usability, subjective comments, and a questionnaire consisting of concerns/interests of drug instructions.

\subsection{Tasks}

\rev{To comprehensively evaluate the effectiveness of the visualization (\emph{Vis}) compared to the traditional text (\emph{Text}) instruction, we designed a multi-faceted assessment. 
This assessment combined objective knowledge tests with subjective usability and cognitive load measures.}
 
The test consisted of 10 single-choice questions designed to assess comprehension of critical information such as indications, dosage, contraindications, and precautions. 
Each correct answer scored 5 points, resulting in total pre- and post-test scores that both ranged from 0 to 50.
We measured performance using two primary metrics: \textbf{``Scores improved''}---the difference between the post-test and pre-test scores (range: 0–50), representing knowledge gain after reading the instructions; \textbf{``Time (s)''}---the time taken (in seconds) to complete the post-test questions, indicating efficiency in locating and understanding information.

To assess subjective experience, we administered:
(1) The \textbf{System Usability Scale (SUS)~\cite{Brooke1996}} to measure usability: by combining different items of the SUS scale~\cite{Brooke1996, Bangor2009, Brooke2013}, we obtained three indicators---learnability, usability, and total usability. All SUS subscales and the total score range from 0 to 100.
(2) The \textbf{NASA Task Load Index (NASA-TLX)~\cite{Hart1988}} to evaluate cognitive load. The NASA-TLX score also ranges from 0 to 100.
(3) A text box to collect \textbf{free-form feedback} on clarity and usefulness.
These measures together allowed us to compare the visualization and text versions across objective performance, efficiency, usability, and cognitive effort.
The complete questionnaires for all tasks, including the knowledge test, SUS, NASA-TLX, and the free-text questions, are provided in the supplemental material (1.supplemental\_material\_main.pdf).

\subsection{Participants}
\rev{
We specifically targeted primary household caregivers (parents or guardians) as participants, as they represent a key demographic of real-world end-users who frequently read OTC drug instructions and make medication decisions for their families. 
To effectively access this group, we recruited on-site outside a school classroom while they waited for their children. 
This setting provided a practical opportunity to engage our target population in a familiar context.
}

\rev{
The entire study was conducted face-to-face by the health data group. 
To ensure a focused evaluation and minimize distractions, the research team designated a quiet area adjacent to, but separate from, the main hallway traffic. 
All interactions and questionnaire completion were carried out in this controlled setting with the researcher's direct supervision, ensuring participant attention remained on the experimental tasks.
}
All participants were parents or guardians of the enrolled students. 

A total of 60 participants over the age of 18 were recruited and randomly assigned to either the Visualization (\emph{Vis}, n=30) or Text (\emph{Text}, n=30) group.
The demographic information of participants is shown in \autoref{tab:basic_info}.
There were no significant differences between the two groups in age (\textbf{$p$}=0.516), gender (\textbf{$p$}=1.000), education level (\textbf{$p$}=0.562), or medical background (\textbf{$p$}=0.278) (\Cref{tab:basic_info}), supporting a valid comparison.
Most of the participants were between 18 and 50 years old (\emph{Vis}: 30, \emph{Text}:28), had obtained higher education (university and above) (\emph{Vis}: 30, \emph{Text}:29), and had no medical background (\emph{Vis}: 27, \emph{Text}:24). 
\rev{They were primary household caregivers for their children, who were likely to read drug instructions and make medication decisions, confirming they matched the profile of our target caregiver user group.}
    
\begin{table}[!htb]
\footnotesize
\centering
\caption{Basic information of participants in the user study.}
\label{tab:basic_info}
\begin{tabular}{lccc}
\toprule
\textbf{Basic information} & \textbf{Vis} & \textbf{Text} & \textbf{$p$-value} \\
\midrule
\textbf{Age (year)} & & & 0.516 \\
18-35 & 9 & 10 & \\
36-50 & 21 & 18 & \\
51-65 & 0 & 1 & \\
66+ & 0 & 1 & \\
\textbf{Gender} & & & 1.000 \\
Male & 11 & 11 & \\
Female & 19 & 19 & \\
\textbf{Education (year)} & & & 0.562 \\
<9 & 0 & 0 & \\
9-12 & 0 & 1 & \\
13-16 & 9 & 10 & \\
>16 & 21 & 19 & \\
\textbf{Medical background} & & & 0.278 \\
Yes & 3 & 6 & \\
No & 27 & 24 & \\
\bottomrule
\end{tabular}
\end{table}
\vspace{-1ex}

\subsection{Procedure}
\label{sec:procedure}

\rev{
The study procedure followed a within-subjects and between-subjects mixed design, comprising four main phases:
(1) background survey and pre-test, (2) intervention and post-test with assigned instruction, (3) subjective rating, and (4) a final comparative feedback session.
}

Laptop computers were used to show the \emph{Vis} version (can switch between the complete and simplified version, with the complete version as the default) or the \emph{Text} version in web browsers.
Participants answered the questionnaire through a survey website (aided by the wjx survey platform~\cite{wjx2024}) 
from their mobile phones.
We briefly introduced to participants the procedure of our experiment and how to use the assigned instruction and the questionnaire before the study began.

All participants were provided informed consent and had to agree to participate. 
After the introduction page, the basic information on age, gender, education level, and medical background was collected through a single-choice question. 
Participants were then invited to indicate their primary concerns or information interests when reading drug instructions through a multiple-choice survey. This question was designed to evaluate whether the content covered in the proposed simplified version aligned with users’ actual priorities. The multiple-choice format, rather than a ranking task, was chosen to efficiently capture the perceived importance of each content item across a list of 15 options while minimizing cognitive burden and better reflecting natural information-seeking behavior.

\textbf{Phase 1: Pre-test.} Next, the actual experiment began, and the participants had to answer 10 knowledge questions regarding paracetamol sustained-release tablets without referring to any instructions. 

\textbf{Phase 2: Intervention \& post-test.} Subsequently, participants were randomly assigned to the \emph{Vis} group of a complete version (experiment) or the \emph{Text} group (control). 
They were asked to answer the 10 questions again while reading their assigned instructions. 
The order of the questions was different for the second pass to avoid the learning effect. 

\textbf{Phase 3: Subjective rating.} Later, they were asked to fill out the SUS and NASA-TLX questionnaires and provide some subjective evaluation/thoughts of their assigned version.
For the \emph{Vis} group, participants were additionally introduced to the simplified version and asked to fill out another SUS after trying it out.
All the information above was documented in the questionnaire.

\textbf{Phase 4: Comparative feedback.} After completing the questionnaire, we presented participants with the drug instructions of the other group, asking for their preferences and feedback using the think-aloud protocol, and recording manually on paper.

The study took an average of 10 minutes, and each participant received compensation equal to the local minimum wage per hour, approximately 3 USD.

\subsection{Data Analysis}

We used the Shapiro-Wilk test to check whether the variables were normally distributed and Levene's test to examine whether the variables satisfied the assumption of homogeneity of variance.
When the variables met the assumptions of normality and homogeneity of variance, the Student’s t-test was used for statistical analysis. Otherwise, the Mann-Whitney U test was used.

\section{Results}
We collected questionnaires from all participants and documented participants’ comments, questions, and general feedback during the evaluation.

\subsection{Visualization (complete) vs. Text}
\begin{figure}[!htb]
  \centering 
  \includegraphics[width=\linewidth]{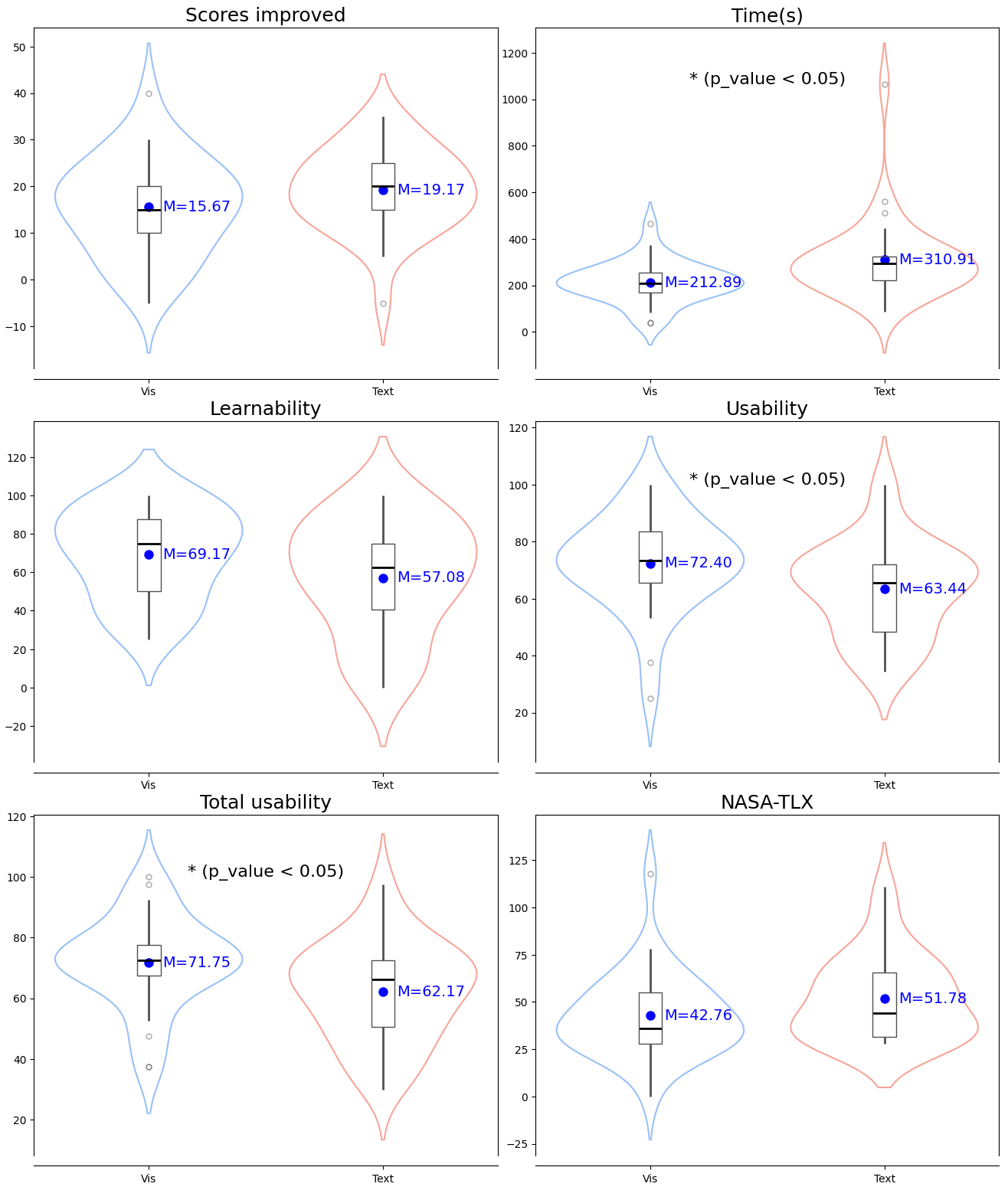}
  \caption{%
  	Results of the user study. Each measurement is visualized with a violin plot showing the data distribution overlaid with a box plot in black. Mean values are shown as blue dots. Statistical differences ($p$\textless0.05) between two groups are marked with ``*''.}
  \label{fig:measurement_violin}
\end{figure}

Distributions of the six indicators of the study results were visualized with violin plots and boxplots as in~\Cref{fig:measurement_violin}.
We did not conduct subgroup analyses, e.g., for age, gender, education level, or medical background, due to insufficient numbers of participants in these subgroups.
``Scores improved” between the groups found no significant difference (\emph{Scores improved}: 15.67$\pm$10.57 vs. 19.17$\pm$8.91, t(58)=-1.387, $p$=0.171). 
The result did not reject nor confirm hypothesis \textbf{H1}.
However, this indicated that our visualization version conveyed drug information with accuracy comparable to the traditional text version, suggesting it served as a complementary alternative to the current version.

The response time of the \emph{Vis} group was significantly shorter than that of the \emph{Text} group (\emph{Time(s)}: 212.89$\pm$90.09 vs. 310.91$\pm$174.25, $z$= 3.001, $p$=\textbf{0.003}), indicating that \textbf{H2} can be accepted. Visualization has the potential to convey information more efficiently. 

The \emph{Vis} version had statistically higher usability rating than that of the \emph{Text} version (\emph{Usability}: 72.4$\pm$16.60 vs. 63.44$\pm$16.56, t(58)=2.093, $p$=\textbf{0.041}; \emph{Total usability}: 71.75$\pm$15.17 vs. 62.17$\pm$16.38, t(58)=2.351, $p$=\textbf{0.022}), indicating that \textbf{H3} can be accepted.

No statistical difference was found for learnability (\emph{Learnability}: 69.17$\pm$23.61 vs. 57.08$\pm$30.21, $z$= -1.487, $p$=0.137) of the two groups. 
However, compared to the \emph{Text} group, the distribution of the \emph{Vis} group was more concentrated and toward high learnability, which indicated easy to learn. 
Further evidence was needed for \textbf{H4} to test the ability of visualization to simplify complex data.

There was no statistically significant difference in NASA-TLX results (\emph{NASA-TLX}: 42.76$\pm$22.46 vs. 51.78$\pm$22.45, $z$= 1.616, $p$=0.106).
However, as shown in~\Cref{fig:measurement_violin}, compared to the \emph{Text} group, the distribution of the \emph{Vis} group was toward lower NASA-TLX, which indicated a lower cognition load. 
Further evidence was needed for \textbf{H5} to test the ability of visualization to reduce information overload. 

Subjective feedback was collected from think-aloud comments and free-text questions. 
Similar responses were grouped, and the frequency of each type of evaluation was summarized in~\Cref{tab:coments}.

\begin{table}[tb]
   \footnotesize
  \centering%
  \caption{Subjective evaluation regarding the visualization and text version instructions.}
  \label{tab:coments}
  \begin{tabular}{%
  	  r%
  	  	*{7}{l}%
  	  	*{2}{r}%
  	}
  	\toprule
  	Group & Comments  &  N\\
  	\midrule
  	Vis &   Clear and easy to understand  &  22\\
  	 &   Key points highlighted  & 10\\
          &   User-friendly & 8\\
          &   Look forward to use  & 4\\
          &   Detailed content   & 3\\
          &   A reasonable distinction of two versions   & 2\\
          &   Easy to download  & 1\\
          &   ``Use with caution'', ``not allowed'' are still &\\
          & difficult to understand  & 2\\
  	Text 
          &   Lack of focus  &  19\\
  	 &   The precautions are unorganized  & 8\\
          &   The content is complex and not intuitive  & 7\\
          &   Need to simplify a little bit  & 6\\
          &   Too professional   & 5\\
          &   Font size is too small  & 4\\
          &   Need instructions for different populations  & 2\\
          &   Need complementary video, animation, pictures   & 2\\  	
  	\bottomrule
  \end{tabular}%
  \\
  \textsuperscript{*} N: the number of comments. 
  \end{table}

Feedback from the visualization group (\emph{Vis}) was predominantly positive. 
High-frequency comments such as ``Clear and easy to understand'' ($N$=22), ``Key points highlighted'' ($N$=10), and ``User-friendly'' ($N$=8) directly reflected the effectiveness of the visual design in enhancing readability and reducing cognitive load. 
Notably, some participants explicitly expressed that they ``look forward to using" ($N$=4) the design, indicating initial practical value and user acceptance.
Regarding the preference for the presentation method of instruction, almost all participants (56 versus 4) chose the visualization version over the traditional text version.
However, a few comments pointed out that icons for precaution legend --- \emph{use with caution} and \emph{not allowed}---remained ``difficult to understand'' ($N$=2). 
This suggested that the semantics of certain icons should be further improved for intuitiveness in the future.

In contrast, feedback from the text group (\emph{Text}) highlighted drawbacks of traditional text-based instructions. 
Users commonly criticized the format for ``Lack of focus'' ($N$=19), ``content is complex and not intuitive'' ($N$=7), and noted that ``The precautions are unorganized'' ($N$=8). 
These critiques underscored the inherent limitations of pure text in information structuring and visual guidance. 
Importantly, some comments directly aligned with the core motivation of our design, such as the need for ``instructions for different populations'' ($N$=2) and for ``complementary video, animation, pictures'' ($N$=2), which from the user's perspective further validated the necessity and significance of developing visualized, tiered instruction design.

In summary, the qualitative user feedback was consistent with the quantitative performance and usability results, collectively demonstrating that the visualization version offered significant advantages in helping users locate information quickly, comprehend key content, and improve the overall user experience. 
The specific comments also provided clear directions for subsequent design refinement.

\subsection{Simplified vs. Complete Vis Versions}
\rev{
To validate the content selection for our two-tiered design (simplified vs. complete), we analyzed data from a survey question administered during the study (see~\Cref{sec:procedure}). 
Participants ($N$=60) were asked to indicate their primary information concerns when reading drug instructions. 
The percentage of participants selecting each section serves as a proxy for its perceived importance to this user group.
}

\begin{figure}[htb]
  \centering 
  \includegraphics[width=0.9\linewidth, trim={1cm, 0.5cm, 0cm, 0cm}]{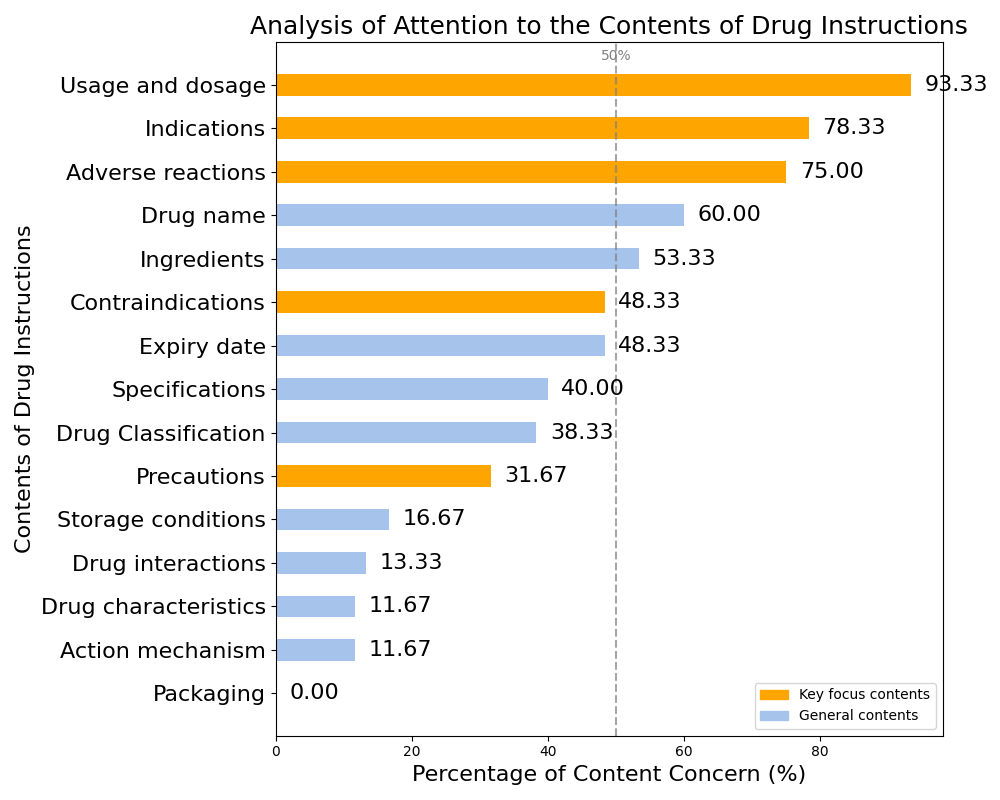}
  \caption{%
  	The main concerns/interests of drug instructions in daily medication were gained from the user study of non-professionals.
    The core sections intentionally emphasized in the simplified design (orange bar) received particularly close attention.}
  \label{fig:content_concern}
\end{figure}

\rev{
The results, ordered by participant concern, are shown in \Cref{fig:content_concern}. 
The core sections intentionally included in the simplified version---indications (78.33\%), usage and dosage (93.33\%), and precautions/contraindications (48.33\%)---were among the highest-ranked concerns. 
In contrast, specialized content (e.g., action mechanism, drug interactions) received lower interest. 
This pattern confirms that the simplified version's content aligns closely with the real-world information needs of the general public for daily medication use.
}

\rev{
While quantitative usability scores showed no significant difference between the simplified (78.15$\pm$11.86) and complete (71.75$\pm$15.17) versions ($t(55)=1.76$, $p=0.084$), qualitative feedback reinforced the two-tiered design's value. 
Although nearly all participants (28/30) agreed the simplified version was sufficient for daily needs, many added they would still consult the complete version for specific, detailed inquiries. 
This supports our design rationale: the simplified version efficiently serves common public needs, while the complete version remains available for deeper information access.
}

\subsection{Workflow for Visualization Design of OTC Drug Instructions}
\begin{figure*}[htb]
  \centering
  \includegraphics[width=0.9\textwidth]{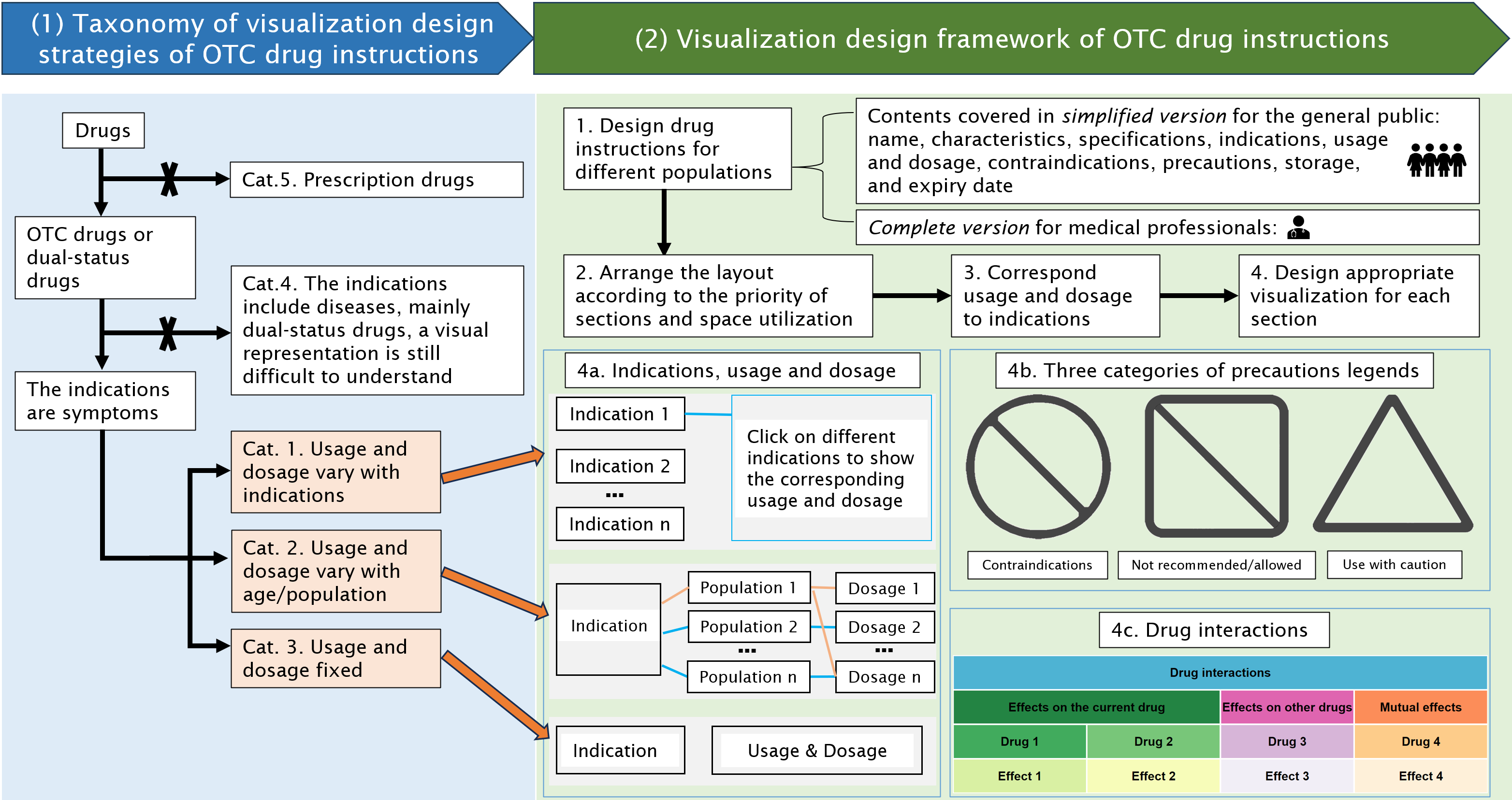}
  \caption{Taxonomy and visualization design framework for OTC drug instructions. (1) Our taxonomy of drug instructions, and in this work, we focus on OTC drug instructions (Categories 1--3, Cat. 1-3). (2) The visualization design framework for OTC drug instructions, which includes four steps (1-4) and specific visualization designs for the most concerning contents (4a-4c). Depending on the category, the specific design of 4a is different.}
  \label{fig:design_otc}
\end{figure*}

\rev{
Integrating the insights from our design study with pharmacists and the empirical results from the user study, we synthesized a systematic workflow for creating visualizations of OTC drug instructions. 
The purpose of this workflow is to provide designers and practitioners with a concrete, step-by-step guide that translates our research findings into actionable practice, ensuring the resulting designs are user-centered and evidence-based. 
The workflow is summarized in \Cref{fig:design_otc}.
}

\rev{The workflow consists of five sequential steps, progressing from categorization to detailed visual encoding:}
\begin{enumerate}
    \item \textbf{Determine the applicable drug category:} Identify if the drug falls into Categories 1–3 of our taxonomy (symptom-based indications) (\Cref{fig:design_otc}(1)). Our framework and validation currently apply to these categories. Drugs for complex conditions (Categories 4–5) may require a distinct design approach.
    \item \textbf{Select and prioritize content for the simplified version:} Base the content selection on empirical user concerns (\Cref{fig:design_otc}(2)-1). In our study, information on usage \& dosage, indications, and precautions (including contraindications and adverse reactions) was rated as most critical by participants. These elements should form the core of the simplified version.
    \item \textbf{Create a layout according to information priority:} Structure the layout to reflect user priorities and efficient visual scanning (\Cref{fig:design_otc}(2)-2). Our findings confirm that placing indications and usage \& dosage most prominently, followed by precautions, aligns with user needs and improves information retrieval, as supported by qualitative feedback and quantitative scores.
    \item \textbf{Map usage and dosage to corresponding indications or populations:} Design a clear visual link (e.g., interactive selection, decision tree, or side-by-side layout as shown in~\Cref{fig:three_dosage_type}) based on the drug's category (\Cref{fig:design_otc}(2)-3). This direct correspondence, validated in our user study, is crucial for preventing dosing errors.
    \item \textbf{Design visual representations grounded in evidence:} Employ visualization types that proved effective in our evaluation (\Cref{fig:design_otc}(2)-4). For example, use hierarchical charts (icicle plots) for drug interactions, traffic-sign-inspired icons for precautions, and highlighted text or icons for key distinctions.
\end{enumerate}
This workflow translates our empirical findings into concrete, actionable steps for creating user-centered OTC drug instructions. While developed for Categories 1–3, its core principles offer a foundation for adapting visualization strategies to other drug information contexts.

\section{Discussion}
In this section, we discuss reflections on study results and visualization design, the limitations of our current work, and potential future directions.

\subsection{Reflections on Study Results and Visualization Design}

While in our user study, there were no significant differences in learnability or NASA-TLX scores between the \emph{Vis} and the \emph{Text} group, the  \emph{Vis} group demonstrated higher usability ratings, faster response time, and more positive user feedback. 
This indicated that our visualization design could contribute to a more user-friendly and easier-to-understand version of drug instructions that holds potential for broader application in the future. 

The analysis of subjective feedback (\Cref{tab:coments}) indicated a trend where participants in the visualization (\emph{Vis}) group provided predominantly positive comments regarding clarity and ease of use, while those in the text (\emph{Text}) group more frequently cited difficulties with information density and navigation. 
This aligned with our quantitative findings of higher usability scores for the \emph{Vis} group. 
However, we acknowledged that participants' awareness of evaluating a new design may have influenced their responses, potentially amplifying criticism of the familiar text format and positive appraisal of the novel visualization.
Therefore, while the qualitative feedback supported the quantitative advantages of the visualization, it should be interpreted in conjunction with the objective performance metrics (task time and accuracy), which were less susceptible to such bias.

The \emph{Vis} group's initial accuracy was slightly higher than that of the \emph{Text} group; still,  both groups demonstrated comparable accuracy after reading the instructions. 
The similar accuracy ceilings across non-professional groups suggested that medical knowledge plays a significant role in comprehension, regardless of the format in which the information was presented. 
Consequently, visualizations of medical content, such as drug instructions, should consider the knowledge level, and it is reasonable to design different visualizations for people with different expertise.
General user interface guidelines stated that content and interaction should be adapted to different target groups and scenarios~\cite{Shneiderman1997, Norman2013TheDO}. 
In our case, both professionals and non-professionals had varying needs for drug information, necessitating tailored designs for specific populations. 
Our current design, which allowed switching between the simplified and complete versions, was effective in addressing these diverse needs. 
The simplified version efficiently delivered essential medication information to the public, ensuring correct use.
At the same time, the complete version, originally intended for medical professionals, also provided detailed information for non-professionals who may require more in-depth knowledge when addressing specific concerns about medications.

Appropriate visualizations for each section of the instructions should be chosen. 
For example, our design employed categorized pictograms to represent varying levels of precaution (see \Cref{fig:design_otc}(2)-4b) and an icicle plot to differentiate hierarchical drug-interaction information (see \Cref{fig:design_otc}(2)-4c).    
It is widely recognized that no single graphic can be universally understood by all individuals~\cite{Burns2022}, as factors including cultural background, age, and education level influence pictogram comprehension~\cite {Abdu-Aguye2023, Chan2013, Beusekom2016}.
In line with this, feedback from our user study indicated that while most precaution icons were readily understood, a few were perceived as overly abstract by some participants. 
This observation underscored the importance of complementing pictograms with concise text labels---a design strategy we implemented. 
Furthermore, to ensure accurate comprehension, our interactive prototype allowed users to access the full, unmodified textual description by hovering over or clicking on any pictogram.
This combined approach---using carefully designed icons supported by immediate textual clarification---proved effective in our evaluation for enhancing information communication. 
It is particularly valuable in cross‑language or low‑literacy scenarios where visuals can provide an accessible entry point to more detailed textual information~\cite{Ng2017}.

\rev{An important design decision involves balancing a clean interface with the discoverability of interactive features.} In the tested prototype, explicit visual affordances (e.g., tooltips, instructional cues, or highlighted interactive areas) were not incorporated to guide users toward interactive elements such as clickable indications or hover-revealed details. 
This design decision was made to prioritize the evaluation of core usability and information comprehension without the influence of prominent guidance elements. To ensure participants were aware of the available interactivity, the researchers provided standardized verbal instructions at the start of the questionnaire, explicitly indicating that hovering or clicking on relevant elements could reveal additional information. This approach allowed participants to begin with equivalent awareness of system functionality while maintaining a clean interface for initial assessment.
We recognize that the absence of built-in visual affordances may limit discoverability in unassisted use, and this will be noted as a consideration for future iterations should the design proceed toward broader deployment.

A web-based format enables our proposed visualization tool to be accessed across various devices (e.g., mobile phones, laptops) without installation.
To bridge physical packaging with digital access, our proposed design can be linked via a QR code printed on the drug package or paper insert, enabling users to instantly access the interactive, visualization-based instructions online---thereby extending rather than replacing the conventional paper format.
Our primary contribution lies not in proposing web or QR access itself---which builds on established practice---but in the design, development, and empirical evaluation of a visualization-based instruction tool tailored for OTC drugs.
While prior research has shown the value of digital health tools (e.g., mobile communication~\cite{Liu2024}, online information~\cite{Lai2024}, digital storytelling~\cite{Yuan2024}), our study provides new evidence that structured visualizations significantly improve comprehension, usability, and informed use compared to traditional text-based instructions. 
The tool is accessible via browser and can also be downloaded or printed for practical reference.

\subsection{Empirical Limitations}
\rev{
While our study provides evidence for the benefits of visualized drug instructions, it is important to acknowledge its limitations regarding expert involvement, participant diversity, design execution, and regulatory status. 
These limitations contextualize the findings and outline areas for future improvement.
}

Due to limited access to a broader pool of professionals, we relied on the same three pharmacists involved in the design process to evaluate the prototype, which may introduce biases such as consensus bias or confirmation bias~\cite{Eastman2022}, and impact the objectivity and generalizability of our findings.
To mitigate these limitations, we conducted two rounds of pilot studies before the main user study. 
Feedback from real users, who are also potential users of the instruction, during these iterations helped refine the prototype and complement the expert evaluations, enhancing the robustness and generalizability of our findings.

In this work, our controlled study primarily involved a relatively young, highly educated population due to practical difficulties in on-site recruitment. Although the two groups (visualization group and text-only group) were comparable in terms of educational background, and most participants were primary family caregivers responsible for medication decisions for children or elderly family members, we recognized that the diversity of the sample in terms of age, health literacy, and other dimensions was limited.
To  evaluate the influence of different user factors on the effectiveness of the design, we expanded the sample size in a follow-up crowd-sourcing study (106 participants in the visualization group and 107 in the text-only control group) and further analyzed the effects of age, gender, educational level, and health literacy on the visualization outcomes.
The results indicated that the visualization design generally offers better usability and lower cognitive load across most user groups, with its benefits being particularly pronounced in non-elderly (under 60 years old), female, university-educated populations, and among individuals with limited health literacy. 
[This study is under review elsewhere.]
The pattern of its advantages varies, suggesting the design's effectiveness is mediated by user characteristics. Further inclusion of more diverse participants would provide more representative evidence for validating and refining the generalizability of the design.

The current study prioritized validating the interactive framework and information hierarchy, which resulted in certain limitations in visual design execution.
Notably, color encodings were used for categorical distinction without clinical semantic meaning, text-background contrast was not fully optimized for accessibility, and icon sets require further refinement for consistency and professionalism. 
These aspects represent clear directions for future work, where adoption of perceptually neutral palettes with clear legends, enhanced contrast for readability, and standardized medical iconography will be essential for real-world deployment and broader user trust.

We conducted a between-subject user study to evaluate the effectiveness of the visualization design but did not account for how keyword search capabilities might affect response time differences between the groups. However, since both groups used a web interface with keyword search functionality on a laptop, the conditions were comparable, and this factor is unlikely to significantly impact the results.

Participants generally found the visualization version of the drug instruction intuitive and usable, but expressed concerns regarding its lack of regulatory approval. 
While expert pharmacists endorsed the design from a professional standpoint, we recommend that---at this stage---the visualized instructions be implemented as a supplementary resource alongside officially approved textual instructions, rather than as a standalone replacement. 
This supplemental approach would allow users to benefit from enhanced comprehension while adhering to current regulatory frameworks. 
In the future, we plan to pursue formal evaluation and approval from relevant health authorities to support the integration of visualization-based instructions into standard medication information materials.

\subsection{Future Work}
\rev{
Several directions for future research are possible, including scaling the design process, extending the framework to more complex drug types, and enhancing it with advanced technologies.
}

Our proposed solution can be extended to other OTC drug instructions by following the established workflow. 
However, scaling this process currently involves considerable manual text processing and design effort. 
To address this, the potential of generative artificial intelligence (AI), such as large language models (LLMs), to semi-automate and accelerate the creation of visual instructions should be explored.
An interactive AI agent that could adapt visual elements based on user proficiency or interest and answer questions about the medication is also desirable.
However, as drug instructions are legal documents critical for safe use, we emphasize that any AI-assisted process must operate under strict human supervision to verify accuracy and prevent potentially harmful misinformation.

In this work, we focused on the design of OTC drug instructions, primarily because non-professionals, including three members of our interdisciplinary research team who do not have medical expertise, face challenges in understanding more complex medical information, such as specific disease indications.
By offering means of simplification as well as details on demand, our framework of OTC drug visualization is beneficial for both non-professional and professional users. 
Pharmacists expressed a strong interest in extending the visualization approach to prescription drug instructions. 
They emphasized the potential benefits of creating customized visualizations tailored to a patient’s specific diagnosis. 
Furthermore, they highlighted the need for professional-level visualizations that incorporate detailed charts, such as pharmacokinetic plots, especially when more comprehensive data becomes available. Future work should explore tools for experts in that regard.

\section{Conclusions}
In this work, we conducted a design study for OTC drug visualization instructions to enhance comprehension.
We systematically searched an official OTC drug database, deriving a taxonomy for the visualization of drug instructions.
Through an iterative design process in collaboration with pharmacists, we developed and implemented visualization versions of instructions for three widely used OTC drugs selected by these experts.
Pharmacist feedback supported our specific design choices, as well as the taxonomy we proposed for OTC drugs.

In addition to expert feedback, we conducted a controlled user study with the general public to evaluate whether our design effectively facilitates information communication. 
The results demonstrated that the visualization version outperformed the traditional text version in terms of response time and usability while maintaining comparable accuracy in understanding the drug paracetamol, and the feedback was more positive.
Our visualization version of drug instructions had the potential to enhance the comprehension of OTC drug information.
The design of the simplified and complete versions met the needs of users in different scenarios.

The design workflow derived from our design study and user study can be potentially extended to the visualization design of many OTC drugs in the future, enhancing people's understanding of various drug information and thereby promoting more rational medication use.


\section*{Declaration of competing interest}
The authors declare that they have no known competing financial interests or personal relationships that could have appeared to influence the work reported in this paper.

\section*{Acknowledgment}
This work was supported by the Beijing Social Science Fund, China 
(grant NO. 25BJ03162).

\section*{Data availability}
Data will be made available on request

 \bibliographystyle{elsarticle-num} 
 \footnotesize
 \bibliography{drugInstruVis}

\end{document}